\documentclass[10pt,journal]{IEEEtran}
\usepackage[latin9]{inputenc}
\usepackage{array}
\usepackage{bm}
\usepackage{multirow}
\usepackage{amsmath}
\usepackage{amssymb}
\usepackage{graphicx}
\usepackage{epsfig}
\usepackage{subfigure}
\usepackage{amsfonts}
\usepackage{cite}
\usepackage{color}
\usepackage{bm}
\usepackage{booktabs}
\usepackage{multirow}
\usepackage{comment}
\usepackage{epstopdf}
\usepackage{stfloats}
\usepackage{makecell}

\providecommand{\tabularnewline}{\\}

\begin{document}
\title{A Full Second-Order Analysis of the Widely Linear MVDR Beamformer for Noncircular Signals}
\author{Zhe Li, \emph{Member},
	\emph{IEEE}, Rui Pu, Yili Xia, \emph{Member},
\emph{IEEE}, Wenjiang Pei, and Danilo P. Mandic, \emph{Fellow},
\emph{IEEE} 
\thanks{Z. Li is with the School of Electronic and Information Engineering, Soochow University, Suzhou 215006, China (e-mail: lizhe@suda.edu.cn).} \thanks{R. Pu, Y. Xia, and W. Pei are with the School of Information Science
and Engineering, Southeast University, 2 Sipailou, Nanjing 210096,
P. R. China. (e-mail: purui@seu.edu.cn;yili\_xia@seu.edu.cn; wjpei@seu.edu.cn).} \thanks{D. P. Mandic is with the Department of Electrical and Electronic Engineering, Imperial College London, London SW7 2AZ, U.K. (e-mail: d.mandic@imperial.ac.uk).}
}
\maketitle
\begin{abstract}
A full performance analysis of the widely linear (WL) minimum variance distortionless response (MVDR) beamformer is introduced. While the WL MVDR is known to outperform its strictly linear counterpart, the Capon beamformer, for noncircular complex signals, the existing approaches provide limited physical insights, since they explicitly or implicitly omit the complementary second-order (SO) statistics of the output interferences and noise (IN). To this end, we exploit the full SO statistics of the output IN to introduce a full SO performance analysis framework for the WL MVDR beamformer. This makes it possible to separate the overall signal-to-interference plus noise ratio (SINR) gain of the WL MVDR beamformer w.r.t. the Capon one into the individual contributions along the in-phase (I) and quadrature (Q) channels. Next, by considering the reception of the unknown signal of interest (SOI) corrupted by an arbitrary number of orthogonal noncircular interferences, we further unveil the distribution of SINR gains in both the I and Q channels, and show that in  almost all the spatial cases, these performance advantages are more pronounced when the SO noncircularity rate of the interferences increases. Illustrative numerical simulations are provided to support the theoretical results.
\end{abstract}
\IEEEpeerreviewmaketitle{ }
\section{Introduction}
\IEEEPARstart{B}{eamforming} is widely recognized as a fundamental technique in array signal processing and its scope of applications includes radar, sonar, wireless communications and spectrum monitoring \cite{Li2005book,Xu2017,Xu2018,Somasundaram2013,Liu2015,Huang2012,Zhuang2016}. The role of beamformers in these applications is to steer the signal of interest (SOI) at particular angles so that the degrading effects of interference can be mitigated and hence the signal quality can be improved. Most conventional beamforming approaches were established on the assumption of stationary observations, which allows the beamformer to be linear and time-invariant. A well-known receiver beamformer in this scenario is the minimum variance distortionless response (MVDR) introduced by Capon in \cite{Capon1969,Capon1967}. The MVDR beamformer minimizes the output power under a linear constraint of the  distortionless SOI, and it is statistically optimal when the covariance matrix and the steering vector of the SOI are known to the receiver. By considering that the available knowledge of the desired signal is often imprecise in practice, several studies have further extended the Capon's method in order to relax its assumptions. Two such robust beamformers were proposed in \cite{Stoica2003} and \cite{Shahbazpanahi2003}, whereby the uncertainties of the SOI array response are modeled through the signal covariance matrix, and via an ellipsoidal set of steering vectors, respectively.

However, such conventional beamformers that build upon linear filters are suboptimal for both nonstationary signals and statistically improper (SO noncircular) signals \cite{Chevalier1996,Chevalier1997}.
Improper signals exhibit different power levels and/or correlation in the real and imaginary channels, and their full second-order (SO) information can be exploited when both the signal itself and its complex conjugate are jointly processed \cite{Picinbono1995}. This, so-called widely linear (WL) processing, has been extensively used in areas including channel equalization \cite{Darsena2005,Mattera2005}, in-phase/quadrature (I/Q) imbalance compensation \cite{Anttila2008,Li2017}, and wireless transmission with improper signaling \cite{Hellings2013,Zeng2013a}, where improper signals appear due to the underlying signal generating physics. For applications in spectrum monitoring and passive listening, where SO noncircular constellations, such as amplitude shift keying, binary phase shift keying, minimum shift keying, and unbalanced quadrature phase shift keying (UQPSK), have been widely used, two types of WL MVDR beamformers were introduced in \cite{Chevalier2007} and in \cite{Chevalier2009}. The former WL MVDR beamformer considers the reception of an unknown signal corrupted by improper interferences, which is SO optimal without requiring any \textit{a priori} knowledge on the SO statistics of the SOI, whereas the latter removes such a limitation by taking into account the SO noncircularity coefficient of both the SOI and interferences, and hence retains its optimality when the unknown SOI exhibits arbitrary noncircularity properties.
Recent efforts have focused on adapting the original WL MVDR beamformer to meet different practical requirements, including reducing computational cost \cite{Liu2018,Shi2015}, specifying SO noncircularity coefficients of input interferences \cite{Xu2013,Xu2014}, and dealing with non-Gaussian and nonlinear improper signals \cite{Huang2016,Chevalier2018,mandic2001}.


Despite the existing extensive applications of WL MVDR beamforming techniques, current theoretical understanding of their operations is still largely based on the pioneering work in \cite{Chevalier2007,Chevalier2009,Chevalier2014}, where their performance gains w.r.t. the Capon beamformer are verified in terms of the signal-to-interference plus noise ratio (SINR), the power ratio between the SOI and the output interferences plus noise (IN). However, from the perspective of augmented complex statistics \cite{Mandic2009,Schreier2010}, these approaches are based on the standard variance analysis of the IN, whereby their complementary SO statistics, a key feature for the processing of improper signals, have been implicitly or explicitly omitted. Motivated by recent advances in augmented complex statistics which have established a novel complementary mean square error analysis to quantify degrees of impropriety of WL estimation errors \cite{Xia20171,Xia2018}, we set out to fill the void in the SINR analysis of the WL MVDR beamformer, in order not only to rigorously analyze their statistical behaviors in a general case (not possible by conventional methods), but also to provide an in-depth characterization of the performance advantage of the WL MVDR beamformer over the Capon one. This makes the proposed analysis suitable for practical beamforming scenarios, e.g., when the antenna array with multiple parallel direct-conversion receivers experiences radio-frequency (RF) I/Q imperfections \cite{Hakkarainen2013,Anttila2008}. The main contributions of this paper are summarized as follows:
%
\begin{itemize}[]
\item We introduce the complementary variance of the output IN, which offers one more degree of freedom to describe the SO behavior of MVDR beamformers. By doing so, the overall SINR performance gain of the WL MVDR beamformer w.r.t. the Capon one can be separated into individual contributions along the I and Q channels, based on the duality of the full SO statistics in the complex domain and the bivariate real domain. This analysis reveals that the SINR gain distribution in the I and Q channels is tightly connected with the overall SINR gain via their respective \textit{distribution coefficients}.
	
\item In order to unveil how the performance advantage offered by the WL MVDR beamformer distributes across the I and Q channels, detailed expressions of the individual SINR gains are derived, covering all the different spatial settings between the SOI and the interferences. The results prove that the performance advantages of the WL MVDR beamformer over the Capon one do exist in both the individual I and Q channels. More specifically, in the most general spatial situation, this makes it possible to elaborate the effect of the SO noncircularity coefficient of the input interferences on the individual SINR gains in the I and Q channels. The so established physical insight shows that the individual gains are both monotonically increasing functions of the SO noncircularity rate of the interferences, whose slopes are related to the SO noncircularity phase of the interferences.

\item For generality, we extend the analysis in \cite{Chevalier2007} by considering the situation where the reception of the SOI is corrupted by an arbitrary number of orthogonal noncircular interferences, instead of the standard two.
\end{itemize}

This paper is organized as follows. Section II briefly introduces basic hypotheses, statistics, and summarizes the formulation of the WL MVDR beamformer, conducted in \cite{Chevalier2007}. After evaluating the individual SINR gains in the I and Q channels based on the full SO statistics of the IN in Section III, Section IV presents a full SINR performance comparison between the WL MVDR beamformer and the Capon one, covering all the spatial settings between the SOI and the SO noncircular interferences. Numerical examples are given in Section V to support the analysis. Finally, Section VI concludes this paper.

\emph{Notations}: Lowercase letters are used to denote scalars, $a$, boldface letters for column vectors, \textbf{a}, and boldface uppercase letters for matrices, \textbf{A}. The symbols ${\textbf 0}_N$, ${\textbf I}_N$, and ${\textbf O}_N$ denote respectively an $N\times 1$ zero vector, an $N\times N$ identity matrix, and an $N\times N$ zero matrix. The superscripts $(\cdot)^{*}$, $(\cdot)^{T}$, $(\cdot)^{H}$ and $(\cdot)^{-1}$ denote respectively the complex conjugation, transpose, Hermitian transpose and matrix inversion operations. The operators $\Re[\cdot]$ and $\Im[\cdot]$ extract respectively the real and imaginary part of a complex variable and $\jmath=\sqrt{-1}$. The statistical expectation operator is denoted by $E[\cdot]$, whose empirical implementation by time-averaging is represented by the operator $\langle \cdot \rangle$.
\section{Hypotheses, Statistics, and Problem Formulation}
\subsection{Hypotheses}
Consider an $N\times1$ complex-valued vector, ${\textbf x}(t)$, which represents the digitized data received from an array consisting of $N$ narrowband sensors at a time instant $t$. Each sensor is supposed to receive the contribution of an SOI corrupted by $P$ statistically uncorrelated far-field narrowband interferences plus background noise, where $P$ is an arbitrary positive integer. In this way, ${{\textbf x}}(t)$ can be represented as
\begin{align}\label{eq:xtdef}
{\textbf x}(t)&\!=\!s(t)e^{\jmath(2 \pi  {f_s}t\!+\! \phi_s)}{{\textbf s}}\!+\!\sum_{p=1}^{P}{m_p}(t)e^{\jmath(2 \pi  {f_p}t+ \phi_p)}{\textbf j}_p\!+\!{\textbf v}(t)\nonumber\\
&\triangleq {s_c}(t)e^{\jmath\phi_s}{\textbf s}\!+\!\sum_{p=1}^{P}m_{cp}(t)e^{\jmath\phi_p}{\textbf j}_p\!+\!{\textbf v}(t)\nonumber\\
&\triangleq {s_c}(t)e^{\jmath\phi_s}{\textbf s}\!+\!{{\textbf v}_T}(t),
\end{align}
where $s(t)$, $ {f_s}$, $\phi_s$ and ${\textbf s}$ respectively denote the complex envelope, the carrier residue, the carrier phase, and the steering vector of the SOI, $s_c(t)\triangleq s(t)e^{\jmath2 \pi  {f_s}t}$. The quantities $m_p(t)$, $ {f_p}$, $\phi_p$ and ${\textbf j}_p$ correspond to the complex envelope, which is potentially SO noncircular, the carrier residue, the carrier phase, and the steering vector of the $p$th interference, $m_{cp}(t)$, to give $m_{cp}(t) = m_p(t)e^{\jmath 2 \pi  {f_p}t}$. The noise vector, ${\textbf v}(t)$, is assumed to be SO circular and stationary with zero-mean, and spatially white. The total noise vector, ${{\textbf v}_T}(t)$, contains both the background noise and the interferences. Note that in \eqref{eq:xtdef} the delay spread in propagation channels is ignored, which is valid for free space propagation and flat fading channels.
%
\subsection{Full SO Statistics}\label{subsec:SOS}
%

Augmented complex statistics have established that the requirement for the description of the full SO statistical behavior of a general complex-valued vector, ${\textbf x}(t)$, is that its covariance matrix, ${\textbf R}_{{\textbf x}}(t,\tau)\triangleq E[{\textbf x}(t){{\textbf x}^H}(t-\tau)]$, and complementary covariance matrix, ${\textbf C}_{{\textbf x}}(t,\tau)\triangleq E[{\textbf x}(t){{\textbf x}^T}(t-\tau)]$, are both employed \cite{Picinbono1995,Schreier2003}. The vector ${\textbf x}(t) $ is said to be SO noncircular if its complementary covariance matrix ${\textbf C}_{{\textbf x}}(t,\tau) \neq {\textbf O}_N$ for at least one couple $(t,\tau)$. When $\tau = 0$, based on \eqref{eq:xtdef}, the covariance matrix ${\textbf R}_{{\textbf x}}$ and the complementary covariance matrix ${\textbf C}_{{\textbf x}}$ of ${\textbf x}(t)$ can be respectively expressed as
\begin{align}
{\textbf R}_{{\textbf x}}& = \langle E[{\textbf x}(t){{\textbf x}^H}(t)] \rangle =\pi_s{\textbf s}{\textbf s}^H+ \sum_{p=1}^{P}\pi_p {\textbf j}_p {\textbf j}_p^H+\eta {\textbf I}_N \nonumber\\&= \pi_s{\textbf s}{\textbf s}^H+{\textbf R},\label{eq:covx}
\end{align}
and
\begin{align}
{\textbf C}_{{\textbf x}}&= \langle E[{\textbf x}(t){{\textbf x}^T}(t)] \rangle =\pi_s \gamma_se^{\jmath 2\phi_s} {\textbf s}{\textbf s}^T+ \sum_{p=1}^{P}\pi_p \gamma_pe^{\jmath 2\phi_p} {\textbf j}_p {\textbf j}_p^T \nonumber\\&=\pi_s \gamma_se^{\jmath 2\phi_s} {\textbf s}{\textbf s}^T+{\textbf C},
\end{align}
where $\pi_s \triangleq \langle E[|s_c(t)|^2] \rangle $ and $\pi_p \triangleq \langle E[|m_{cp}(t)|^2] \rangle $ are the time-averaged powers of the SOI and the $p$th interference received by an omnidirectional sensor, respectively, {$\eta$ is the mean power of the noise per sensor}, $\gamma_s \triangleq \langle E[s_c^2(t)] \rangle / \pi_s$ and $\gamma_p \triangleq \langle E[m_{cp}^2(t)] \rangle / \pi_p$ are the time-averaged SO noncircularity coefficients of the SOI and the $p$th interference, respectively. Moreover, $\gamma_s=|\gamma_s|e^{\jmath\delta_s}$ and $\gamma_p=|\gamma_p|e^{\jmath\delta_p}$, where $|\gamma_s|$ and $|\gamma_p|$ denote their time-averaged SO noncircularity rates, while ${\delta_s}$, and $\delta_p$ represent their time-averaged SO noncircularity phases and ${\textbf R} \triangleq \langle E[{\textbf v}_T(t){\textbf v}_T^H(t)] \rangle$ and  ${\textbf C} \triangleq \langle E[{\textbf v}_T(t){\textbf v}_T^T(t)] \rangle$ are the sample covariance and complementary covariance matrix of the total noise, ${\textbf v}_T(t)$, when $\tau=0$.

\subsection{WL MVDR Beamformer}
For the reception of an unknown SOI, corrupted by potentially SO noncircular interferences, the optimal WL MVDR beamformer uses a $2N \times 1$ WL spatial filter $\widetilde{{\bm \omega}}\triangleq [{\bm \omega}_1^T,{\bm \omega}_2^T]^T$ to yield an output, $y(t)$, given by \cite{Chevalier2007}
\begin{equation}\label{eq:yt}
y(t)\triangleq {\bm \omega}_1^H {\textbf x}(t)+{\bm \omega}_2^H {{\textbf x}^\ast}(t) \triangleq \widetilde{{\bm \omega}}^H \widetilde{{\textbf x}}(t),
\end{equation}
where $\widetilde{{\textbf x}}(t) \triangleq [{\textbf x}^T(t),{\textbf x}^H(t)]^T$ is the augmented observation vector. Upon inserting \eqref{eq:xtdef} into \eqref{eq:yt}, the output $y(t)$ can be expressed as
\begin{align}\label{eq:yt2}
y(t)=s_c(t)e^{\jmath \phi_s} \widetilde{{\bm \omega}}^H \widetilde{{\textbf s}}_1 + {s_c}^{\ast}(t)e^{-\jmath \phi_s} \widetilde{{\bm \omega}}^H \widetilde{{\textbf s}}_2 + \widetilde{{\bm \omega}}^H \widetilde{{\textbf v}}_T(t),
\end{align}
where $\widetilde{{\textbf s}}_1 \triangleq [{\textbf s}^T,{\bf 0}_N^T]^T$, $\widetilde{{\textbf s}}_2 \triangleq [{\bf 0}_N^T,{\textbf s}^H]^T$ and $\widetilde{{\textbf v}}_T(t) \triangleq [{\textbf v}_T^T(t),{\textbf v}_T^H(t)]^T$.

It is important to notice that finding the optimal filter coefficient vector ${\bm \omega}$ which generates the best estimate of the SOI, $s_c(t)e^{\jmath \phi_s}$, when its time-averaged SO noncircularity coefficient, $\gamma_s$, is unknown, is equivalent to generating the output $y(t)$ by solving the constrained optimization problem based on the non-null SOI, $s_c(t)$, and its complex conjugate $s_c^*(t)$, given by
\begin{align}
\min& \quad \widetilde{{\bm \omega}}^H{\textbf R}_{\widetilde{{\textbf x}}}\widetilde{{\bm \omega}} \nonumber\\
{\rm s.t.}& \quad \widetilde{{\bm \omega}}^H \widetilde{{\textbf s}}_1=1 \ \text{and} \ \ \widetilde{{\bm \omega}}^H \widetilde{{\textbf s}}_2=0,\label{eq:WLMVDROp}
\end{align}
where ${\textbf R}_{\widetilde{{\textbf x}}} \triangleq \langle E[\widetilde{{\textbf x}}(t) {\widetilde{{\textbf x}}}^H(t)] \rangle$ is the time-averaged covariance matrix of the augmented observation $\widetilde{{\textbf x}}(t)$. The solution to this problem is given by
\begin{align}\label{eq:wMVDR}
\widetilde{{\bm \omega}}_{\scriptscriptstyle {\text {MVDR}}} ={\textbf R}_{\widetilde{{\textbf v}}}^{-1} {\bf S} [{\bf S}^H {\textbf R}_{\widetilde{{\textbf v}}}^{-1} {\bf S}]^{-1} \bf{f},
\end{align}
where ${\bf S} \triangleq [\widetilde{{\textbf s}}_1,\widetilde{{\textbf s}}_2]$ is a $2N \times 2$ matrix composed by the steering vector ${\textbf s}$ and ${\bf 0}_N$, ${\bf f} \triangleq [1,0]^T$ is a $2 \times 1$ constant vector, and ${\textbf R}_{\widetilde{{\textbf v}}} \triangleq \langle E[\widetilde{{\textbf v}}_T(t) \widetilde{{\textbf v}}_T^H(t)] \rangle$ is the $2N \times 2N$ time-averaged augmented covariance matrix of the total noise ${\textbf v}_T(t)$, which can be further decomposed as
\begin{align}\label{eq:acovRv}
{\textbf R}_{\widetilde{{\textbf v}}}=
\left[ {\begin{array}{*{20}{c}}
{\textbf R} & {\textbf C} \\ {\textbf C}^\ast & {\textbf R}^\ast
\end{array}} \right].
\end{align}
Upon substituting ${\bm \omega}_{\scriptscriptstyle {\text {MVDR}}}$ in \eqref{eq:wMVDR} into \eqref{eq:yt2}, the overall output, $y(t)$, can be expressed as
\begin{equation}\label{eq:outputyt}
y(t)\triangleq s_c(t)e^{\jmath \phi_s}+q_{\scriptscriptstyle {\text {MVDR}}}(t),
\end{equation}
where
\begin{equation}\label{eq:qMVDRt}
q_{\scriptscriptstyle {\text {MVDR}}}(t)={\bf{f}}^H [{\bf S}^H {\textbf R}_{\widetilde{{\textbf v}}}^{-1} {\bf S}]^{-1} {\bf S}^H {\textbf R}_{\widetilde{{\textbf v}}}^{-1} {\widetilde{\textbf v}}_T(t),
\end{equation}
is the output IN.

Note that the Capon MVDR beamformer \cite{Capon1969,Capon1967}
%
\begin{equation}\label{eq:Capon}
{\bm \omega}_{\scriptscriptstyle {\text {Capon}}} = ({\textbf s}^H {\textbf R}^{-1} {\textbf s})^{-1} {{\textbf R}}^{-1} {\textbf s},
\end{equation}
is the solution to the strictly linear constrained optimization problem, given by
\begin{align}
\min& \quad {\bm \omega}^H{\textbf R}_{{\textbf x}}{\bm \omega} \nonumber\\
{\rm s.t.}& \quad {\bm \omega}^H {\textbf s}=1,
\end{align}
which indicates that ${\bm \omega}_{\scriptscriptstyle {\text {Capon}}}$  can be regarded as a reduced version of the WL MVDR beamformer $\widetilde{{\bm \omega}}_{\scriptscriptstyle {\text {MVDR}}}$ in \eqref{eq:wMVDR}, by considering the steering vector of the SOI ${\textbf s}$ itself instead of its augmented matrix ${\bf S}$. In general, the Capon beamformer, ${\bm \omega}_{\scriptscriptstyle {\text {Capon}}}$, in \eqref{eq:Capon}, is suboptimal for the augmented optimization problem in \eqref{eq:WLMVDROp}, and is equivalent to the optimal WL MVDR beamformer $\widetilde{{\bm \omega}}_{\scriptscriptstyle {\text {MVDR}}}$ only when the total noise, ${\textbf v}_T(t)$, is SO circular with a vanishing complementary covariance matrix ${\textbf C}={\bf O}_N$. Correspondingly, its output IN $q_{\scriptscriptstyle{\text{Capon}}}(t)$ can be derived from \eqref{eq:qMVDRt} as
\begin{equation}\label{eq:qCapon}
q_{\scriptscriptstyle{\text {Capon}}}(t) = ({\textbf s}^H {\textbf R}^{-1} {\textbf s})^{-1}{\textbf s}^H {{\textbf R}}^{-1}{\textbf v}_T(t).
\end{equation}
\section{Individual SINR gains in the I and Q Channels}\label{sec:defs}
The conventional metric to evaluate the performance of a beamformer is the signal-to-interference plus noise ratio (SINR), which is defined as the power ratio between the SOI and the output IN. According to the analysis in \cite{Chevalier2007}, the SINR of the WL MVDR beamformer $\widetilde{{\bm \omega}}_{\scriptscriptstyle {\text {MVDR}}}$ can be evaluated as
%
\begin{align}\label{eq:SINRdef}
\text{SINR}_{\scriptscriptstyle {\text {MVDR}}} = \frac{\pi_s}{\kappa_{\scriptscriptstyle {\text {MVDR}}}},
\end{align}
where $\kappa_{\scriptscriptstyle {\text {MVDR}}}$ denotes the time-averaged variance of the output IN $q_{\scriptscriptstyle {\text {MVDR}}}(t)$, given by
\begin{align}\label{eq:kappa}
\kappa_{\scriptscriptstyle {\text {MVDR}}} \triangleq \langle E[|q_{\scriptscriptstyle {\text {MVDR}}}(t)|^2] \rangle = {\bf{f}}^H [{\bf S}^H {\textbf R}_{\widetilde{{\textbf v}}}^{-1} {\bf S}]^{-1} \bf{f}.
\end{align}

An inspection of \eqref{eq:SINRdef} and \eqref{eq:kappa} illustrates that the performance of the WL MVDR beamformer is dependent on the augmented covariance matrix, ${\textbf R}_{\widetilde{{\textbf v}}}$, which represents the full SO statistics of the complex-valued total noise vector, ${\textbf v}_T(t)$. However, as discussed in Section \ref{subsec:SOS}, from the perspective of augmented complex statistics \cite{Mandic2009,Schreier2010}, the SINR metric in \eqref{eq:SINRdef} utilizes partial statistical information of the noncircular output IN $q_{\scriptscriptstyle {\text {MVDR}}}(t)$, as its complementary variance is omitted. Since the primary goal of the WL MVDR beamformer is to employ a WL signal processing framework to deal with the SO noncircular IN, it is a prerequisite to investigate how the input noncircularity propagates into the WL MVDR beamformer. To address this issue, we first explore the full SO statistics of the output IN. Next, based on the duality between the complex domain and the bivariate real domain, we further provide an in-depth characterization on how its SINR gain is distributed across physical I and Q channels, and finally benchmark the results against the Capon SINR.
%
%
\subsection{Full SO Statistics of the Output IN}\label{subsec_SOS2}
The complementary variance of the output IN, $q_{\scriptscriptstyle {\text {MVDR}}}(t)$, can be defined as  $\widetilde{\kappa}_{\scriptscriptstyle {\text {MVDR}}} \triangleq \langle E[q_{\scriptscriptstyle {\text {MVDR}}}(t)q_{\scriptscriptstyle {\text {MVDR}}}(t)] \rangle$. According to \eqref{eq:qMVDRt}, $\widetilde{\kappa}_{\scriptscriptstyle {\text {MVDR}}}$ can be further expressed as
\begin{equation}\label{eq:tildekappa}
\widetilde{\kappa}_{\scriptscriptstyle {\text {MVDR}}} \!=\! {\bf{f}}^H \![{\bf S}^H {\textbf R}_{\widetilde{{\textbf v}}}^{-1} {\bf S}]^{-1} [{\bf S}^H ({{\textbf C}_{\widetilde{{\textbf v}}}^{-1}})^* {\bf S}^\ast]^T\! ([{\bf S}^H {\textbf R}_{\widetilde{{\textbf v}}}^{-1} {\bf S}]^{-1})^T \bf{f}^*\!,
\end{equation}
where ${\textbf C}_{\widetilde{{\textbf v}}} \triangleq \langle E[\widetilde{{\textbf v}}_T(t) \widetilde{{\textbf v}}_T^T(t)] \rangle$ is the sample augmented complementary covariance matrix of the total noise, ${{\textbf v}}_T(t)$, which can be partitioned into a block matrix as \cite{Mandic2015}
\begin{align}\label{eq:apseucovRv}
{\textbf C}_{\widetilde{{\textbf v}}}=
\left[ {\begin{array}{*{20}{c}}
{\textbf C} & {\textbf R} \\ {\textbf R}^\ast & {\textbf C}^\ast
\end{array}} \right].
\end{align}
Using the matrix inversion lemma, from  \eqref{eq:acovRv} and \eqref{eq:apseucovRv} we have
\begin{equation}\label{eq:invR}
{\textbf R}_{\widetilde{{\textbf v}}}^{-1}=
\left[ {\begin{array}{*{20}{c}}
\bf{A} & \bf{D} \\ \bf{D}^\ast & \bf{A}^\ast
\end{array}} \right],
\end{equation}
\begin{equation}\label{eq:invC}
{\textbf C}_{\widetilde{{\textbf v}}}^{-1}=
\left[ {\begin{array}{*{20}{c}}
\bf{D}^\ast & \bf{A}^\ast \\ \bf{A} & \bf{D}
\end{array}} \right],
\end{equation}
where the $N \times N$ Hermitian matrix $\bf{A}$ and the $N \times N$ symmetric matrix $\bf{D}$ are defined as
\begin{equation}\label{eq:AD}
{\bf{A}} \!\triangleq\! {[{\textbf R}-{\textbf C} {\textbf R}^{* -1} {\textbf C}^\ast]}^{-1}, ~ {\bf D} \!\triangleq\! -{[{\textbf R}-{\textbf C} {\textbf R}^{*-1} {\textbf C}^\ast]}^{-1} {\textbf C} {\textbf R}^{*-1}.
\end{equation}
Next, upon substituting \eqref{eq:invR} and \eqref{eq:invC} into \eqref{eq:kappa} and \eqref{eq:tildekappa}, we have
\begin{eqnarray}
\kappa_{\scriptscriptstyle {\text {MVDR}}} \!\!\!\!&=&\!\!\!\! \frac{{\textbf s}^H \bf{A} {\textbf s}}{{|{\textbf s}^H \bf{A} {\textbf s}|}^2-{|{\textbf s}^H \bf{D} {\textbf s}^\ast|}^2},\label{eq:kappaexplicit}\\
\widetilde{\kappa}_{\scriptscriptstyle {\text {MVDR}}} \!\!\!\!&=&\!\!\!\! -\frac{{\textbf s}^H \bf{D} {\textbf s}^\ast}{{|{\textbf s}^H \bf{A} {\textbf s}|}^2-{|{\textbf s}^H \bf{D} {\textbf s}^\ast|}^2}.\label{eq:tildekappaexplicit}
\end{eqnarray}
For comparison, the output SINR of the standard Capon beamformer is given by \cite{Applebaum1976}
\begin{equation}\label{eq:SINRCapon1}
{\text{SINR}_{\scriptscriptstyle {\text {Capon}}}}= \pi_s {\textbf s}^H {\textbf R}^{-1} {\textbf s},
\end{equation}
whose sample standard and complementary variance of the output IN, $\kappa_{\scriptscriptstyle {\text {Capon}}}$ and ${\widetilde{\kappa}}_{\scriptscriptstyle {\text {Capon}}}$, can be respectively derived as
\begin{align} \label{eq:kappa_V}
\kappa_{\scriptscriptstyle {\text {Capon}}} \triangleq \langle E[|q_{\scriptscriptstyle {\text {Capon}}}(t)|^2] \rangle =({\textbf s}^H {\textbf R}^{-1} {\textbf s})^{-1},
\end{align}
and
\begin{align} \label{eq:tildekappa_V}
{\widetilde{\kappa}}_{\scriptscriptstyle {\text {Capon}}} \triangleq \langle E[q_{\scriptscriptstyle {\text {Capon}}}(t)q_{\scriptscriptstyle {\text {Capon}}}(t)] \rangle= \frac{{\textbf s}^H {\bf R}^{-1}{\bf C}{\bf R}^{\ast -1} {\textbf s}^\ast}{|{\textbf s}^H {\bf R}^{-1} {\textbf s}|^2}.
\end{align}

In a particular case when the total noise, ${\textbf v}_T(t)$, is SO circular, i.e., its complementary covariance matrix  becomes ${\textbf C}={\bf O}_N$, which also leads to ${\textbf D}={\bf O}_N$, based on \eqref{eq:kappaexplicit}, \eqref{eq:tildekappaexplicit}, \eqref{eq:kappa_V} and \eqref{eq:tildekappa_V}, we have $\widetilde{\kappa}_{\scriptscriptstyle {\text {MVDR}}}=\widetilde{\kappa}_{\scriptscriptstyle {\text {Capon}}}=0$, and $\kappa_{\scriptscriptstyle {\text {MVDR}}}=\kappa_{\scriptscriptstyle {\text {Capon}}}$, which indicates that both the WL MVDR and Capon beamformers are identical, and confirms the generality of our approach.
%
\subsection{Individual SINR Gains in the I and Q Channels}
The standard SINR analysis in \cite{Chevalier2007} gives the SINR gain, $G$, as
\begin{equation}\label{eq:G_def}
G \triangleq \frac{{\text{SINR}}_{\scriptscriptstyle {\text {MVDR}}}}{\text{SINR}_{\scriptscriptstyle {\text {Capon}}}}=\frac{\kappa_{\scriptscriptstyle {\text {Capon}}}}{\kappa_{\scriptscriptstyle {\text {MVDR}}}}
\end{equation}
which reflects only the output error power difference between the two beamformers, and is not sufficient to model its distribution across I and Q channels. On the other hand, the complementary SO statistics of the output IN, that is, $\widetilde{\kappa}_{\scriptscriptstyle {\text {MVDR}}}$ and $\widetilde{\kappa}_{\scriptscriptstyle {\text {Capon}}}$, evaluated in Section \ref{subsec_SOS2}, offers an extra degree of freedom to describe the SO statistics in the real and imaginary channels. To illustrate this modeling advantage, we first decompose the output IN of the WL MVDR beamformer, $q_{\scriptscriptstyle {\text {MVDR}}}(t)$, into its respective real and imaginary parts as
\begin{equation}\label{eq:qtReIm}
q_{\scriptscriptstyle {\text {MVDR}}}(t) = \Re[q_{\scriptscriptstyle {\text {MVDR}}}(t)]+\jmath \Im[q_{\scriptscriptstyle {\text {MVDR}}}(t)].
\end{equation}
Then, upon introducing $(\kappa_{\scriptscriptstyle {\text {MVDR}}})_I\triangleq \langle E\left[\{\Re[q_{\scriptscriptstyle {\text {MVDR}}}(t)]\}^2\right] \rangle$ and $(\kappa_{\scriptscriptstyle {\text {MVDR}}})_Q\triangleq \langle E\left[\{\Im[q_{\scriptscriptstyle {\text {MVDR}}}(t)]\}^2\right] \rangle$ as the time-averaged power of the output IN in the I and Q channels, respectively, and based on the definition of $\kappa_{\scriptscriptstyle {\text {MVDR}}} $ in \eqref{eq:kappa} and $\widetilde{\kappa}_{\scriptscriptstyle {\text {MVDR}}}$ in \eqref{eq:tildekappa}, we have
\begin{align}
(\kappa_{\scriptscriptstyle {\text {MVDR}}})_I\triangleq\langle E\left[\{\Re[q_{\scriptscriptstyle {\text {MVDR}}}(t)]\}^2\right] \rangle&=\frac{1}{2} (\kappa_{\scriptscriptstyle {\text {MVDR}}}+ \Re[\widetilde{\kappa}_{\scriptscriptstyle {\text {MVDR}}}]) \label{eq:kappaI},
\\(\kappa_{\scriptscriptstyle {\text {MVDR}}})_Q\triangleq \langle E\left[\{\Im[q_{\scriptscriptstyle {\text {MVDR}}}(t)]\}^2\right] \rangle&=\frac{1}{2} (\kappa_{\scriptscriptstyle {\text {MVDR}}}- \Re[\widetilde{\kappa}_{\scriptscriptstyle {\text {MVDR}}}]). \label{eq:kappaQ}
\end{align}
Similarly, for the Capon beamformer we can also separate its output error power $\kappa_{\scriptscriptstyle {\text {Capon}}}$ into I and Q channels as
\begin{align}
(\kappa_{\scriptscriptstyle {\text {Capon}}})_I\triangleq\langle E\left[\{\Re[q_{\scriptscriptstyle {\text {Capon}}}(t)]\}^2\right] \rangle&=\frac{1}{2} (\kappa_{\scriptscriptstyle {\text {Capon}}}+ \Re[\widetilde{\kappa}_{\scriptscriptstyle {\text {Capon}}}]) \label{eq:kappaICapon},
\\(\kappa_{\scriptscriptstyle {\text {Capon}}})_Q\triangleq \langle E\left[\{\Im[q_{\scriptscriptstyle {\text {Capon}}}(t)]\}^2\right] \rangle&=\frac{1}{2} (\kappa_{\scriptscriptstyle {\text {Capon}}}- \Re[\widetilde{\kappa}_{\scriptscriptstyle {\text {Capon}}}]). \label{eq:kappaQCapon}
\end{align}

Now, a joint consideration of the standard and complementary variances of the output INs in \eqref{eq:kappaI}-\eqref{eq:kappaQCapon} precisely quantifies the individual performance advantages of the WL MVDR beamformer over the Capon one in both the I and Q channels by considering
\begin{equation}\label{eq:GIdef}
G_I \triangleq \frac{(\kappa_{\scriptscriptstyle {\text {Capon}}})_I}{(\kappa_{\scriptscriptstyle {\text {MVDR}}})_I},
\end{equation}
and
\begin{equation}\label{eq:GQdef}
G_Q \triangleq \frac{(\kappa_{\scriptscriptstyle {\text {Capon}}})_Q}{(\kappa_{\scriptscriptstyle {\text {MVDR}}})_Q}.
\end{equation}
Upon inserting \eqref{eq:kappaI} and \eqref{eq:kappaICapon} into \eqref{eq:GIdef}, we have
\begin{align}\label{eq:GIonG}
G_I &=\frac{\kappa_{\scriptscriptstyle {\text {Capon}}}+ \Re[\widetilde{\kappa}_{\scriptscriptstyle {\text {Capon}}}]}{\kappa_{\scriptscriptstyle {\text {MVDR}}}+ \Re[\widetilde{\kappa}_{\scriptscriptstyle {\text {MVDR}}}]}\nonumber\\
&= \frac{(\kappa_{\scriptscriptstyle {\text {Capon}}}+ \Re[\widetilde{\kappa}_{\scriptscriptstyle {\text {Capon}}}])/\kappa_{\scriptscriptstyle {\text {Capon}}}}{(\kappa_{\scriptscriptstyle {\text {MVDR}}}+ \Re[\widetilde{\kappa}_{\scriptscriptstyle {\text {MVDR}}}])/\kappa_{\scriptscriptstyle {\text {MVDR}}}} \cdot \frac{\kappa_{\scriptscriptstyle {\text {Capon}}}}{\kappa_{\scriptscriptstyle {\text {MVDR}}}}\nonumber\\
&= \frac{1+\Re[\gamma_{q,{\scriptscriptstyle {\text {Capon}}}}]}{1+\Re[\gamma_{q,{\scriptscriptstyle {\text {MVDR}}}}]} \cdot G,
\end{align}
where
\begin{equation}\label{eq:gammaqdef}
\gamma_{q,{\scriptscriptstyle {\text {Capon}}}} \triangleq  \frac{\widetilde{\kappa}_{\scriptscriptstyle {\text {Capon}}}}{\kappa_{\scriptscriptstyle {\text {Capon}}}} \ \text{and} \ \gamma_{q,{\scriptscriptstyle {\text {MVDR}}}} \triangleq \frac{\widetilde{\kappa}_{\scriptscriptstyle {\text {MVDR}}}}{\kappa_{\scriptscriptstyle {\text {MVDR}}}},
\end{equation}
are the time-averaged SO noncircularity coefficients of the output IN of the Capon beamformer, $q_{\scriptscriptstyle {\text {Capon}}}(t)$, and the WL MVDR beamformer, $q_{\scriptscriptstyle {\text {MVDR}}}(t)$, respectively. Based on \eqref{eq:kappaexplicit}-\eqref{eq:tildekappa_V}, they can be further evaluated as
\begin{equation}\label{eq:gammaqderv}
\gamma_{q,{\scriptscriptstyle {\text {Capon}}}} = \frac{{\textbf s}^H {\bf R}^{-1}{\bf C}{\bf R}^{\ast -1} {\textbf s}^\ast}{{\textbf s}^H {\bf R}^{-1} {\textbf s}} \ \text{and} \ \gamma_{q,{\scriptscriptstyle {\text {MVDR}}}} =-\frac{{\textbf s}^H \bf{D} {\textbf s}^\ast}{{\textbf s}^H \bf{A} {\textbf s}}.
\end{equation}
Similar to \eqref{eq:GIonG}, by inserting \eqref{eq:kappaQ} and \eqref{eq:kappaQCapon} into \eqref{eq:GQdef}, we obtain
\begin{equation}\label{eq:GQonG}
G_Q = \frac{1-\Re[\gamma_{q,{\scriptscriptstyle {\text {Capon}}}}]}{1-\Re[\gamma_{q,{\scriptscriptstyle {\text {MVDR}}}}]} \cdot G.
\end{equation}
Now, based on \eqref{eq:GIonG} and \eqref{eq:GQonG}, we can further define the \textit{distribution coefficients}, $\lambda_I$ and $\lambda_Q$, as
\begin{align} \label{eq:dis_co}
\lambda_I \triangleq \frac{1+\Re[\gamma_{q,{\scriptscriptstyle {\text {Capon}}}}]}{1+\Re[\gamma_{q,{\scriptscriptstyle {\text {MVDR}}}}]}  ~\text{and}~\lambda_Q \triangleq \frac{1-\Re[\gamma_{q,{\scriptscriptstyle {\text {Capon}}}}]}{1-\Re[\gamma_{q,{\scriptscriptstyle {\text {MVDR}}}}]},
\end{align}
so that individual SINR gains in the I and Q channels are associated with the overall SINR gain $G$ through their respective distribution coefficients, that is,
\begin{equation}
G_I = \lambda_I \cdot G \ \ \text{and} \ \ G_Q = \lambda_Q \cdot G. \label{eq:dis_G}
\end{equation}

\textit{Remark 1:}
Expression \eqref{eq:dis_G} reveals that the individual SINR gains in the I and Q channels are both proportional to the overall gain. Their respective  distribution coefficients, $\lambda_I$ and $\lambda_Q$ are related with SO noncircularity coefficients of output INs of the WL MVDR and the Capon beamformers, that is, $\gamma_{q,{\scriptscriptstyle {\text {MVDR}}}}$ and $\gamma_{q,{\scriptscriptstyle {\text {Capon}}}}$. By definition, we have  $-1<\Re[q_{\scriptscriptstyle {\text {Capon}}}(t)]<1$ and $-1<\Re[q_{\scriptscriptstyle {\text {MVDR}}}(t)]<1$, and hence, based on \eqref{eq:dis_co}, the distribution coefficients are strictly positive, i.e., $0<\lambda_I<\infty$ and $0<\lambda_Q<\infty$.
\section{A FULL SINR Gain Analysis}\label{Sec:SINRGainAna}
It is of particular interest to find an explicit link between the SO noncircularity coefficient of the input interferences, so as to demonstrate the advantages of the WL MVDR beamformer over the Capon one in detail. For generality, we consider the situation where the reception of the SOI is corrupted by an arbitrary number of noncircular interferences. For ease of the analysis, physical meanings of the symbols defined in this section are summarized in Table I.

%
In order to interpret the relationship among the SOI, the $P$ interferences and the noise within the observed signal ${\textbf x}(t)$, we define
\begin{equation}
\varepsilon_s \triangleq ({\textbf s}^H {\textbf s}) \pi_s / \eta \hspace{50pt} \label{power_s}
\end{equation}
as the power ratio between the SOI and the noise,
\begin{equation}
\varepsilon_p \triangleq ({\textbf j}_p^H {\textbf j}_p) \pi_p / \eta \hspace{46pt} \label{eq:power_p}
\end{equation}
as the power ratio between the $p$th interference and the noise, where $p \in \{1, 2, \ldots,P\}$, and
\begin{equation}
\alpha_{ps} \triangleq \frac{{\textbf j}_p^H {\textbf s}}{({\textbf j}_p^H {\textbf j}_p)^{1/2} ({\textbf s}^H {\textbf s})^{1/2}} \triangleq |\alpha_{ps}|e^{\jmath \phi_{ps}} \label{eq:spatial}
\end{equation}
as the spatial correlation coefficient between the $p$th interference and the SOI,  such that $0 \leq |\alpha_{ps}| \leq 1$.

For ease of computation, we follow the analysis in \cite{Chevalier2007} to give specific constraints on the interferences:

%
\textit{Assumption 1:} All the input interferences have uniform power and SO noncircularity coefficient, that is, $\forall p \in \{1, 2, \ldots,P\}$, $\pi_p = \pi$, such that $\varepsilon_p = \varepsilon$ according to \eqref{eq:power_p}, and $ |\gamma_p|e^{\jmath\delta_p}=|\gamma|e^{\jmath\delta}$, where $0<|\gamma|<1$.

\textit{Assumption 2:} The steering vectors of the input interferences are orthogonal to each other, i.e., $\forall  i,j \in \{1,\ldots,P\}$, and $i\neq j$, ${\textbf j}_i^H {\textbf j}_j=0$.

The above constraints statistically describe a general situation where the SOI is received by a linear array of uniform sensors spaced half a wavelength apart, and is corrupted by $P$ SO noncircular interferences that are modulated with the same improper constellation.

%
%
\begin{table}[t!]
		\centering
		\caption{Notations and symbols defined in Section IV.}
		\label{Tab:denotation} %
		\begin{tabular}{|c|l|}
			\hline
			Symbol & \multicolumn{1}{c}{Denotation}\tabularnewline
			\hline
			$\varepsilon_s$ & The power ratio between the SOI and the noise\tabularnewline		
			\hline
			$\varepsilon_p$ & The power ratio between the $p$th interference and the noise\tabularnewline	
			\hline
			$\varepsilon$ & \begin{tabular}{@{}l@{}}The power ratio between the interference and the noise under \\ \textit{Assumption 1}\end{tabular}  \tabularnewline	
			\hline
			$\pi$ & The power of the interference under \textit{Assumption 1} \tabularnewline		
			\hline
			$\alpha_{ps}$ &
			\begin{tabular}{@{}l@{}}The spatial correlation coefficient between the $p$th \\ interference and the SOI\end{tabular}\tabularnewline		
			\hline
			$\phi_{ps}$ & The phase of the coefficient $\alpha_{ps}$\tabularnewline		
			\hline
			$\gamma$ & \begin{tabular}{@{}l@{}}The SO noncircularity coefficient of the interference under \\ \textit{Assumption 1}\end{tabular}\tabularnewline
			\hline					
			$\delta$ & \begin{tabular}{@{}l@{}}The SO noncircularity phase of the interference under \\ \textit{Assumption 1}\end{tabular} \tabularnewline	
			\hline
			$|\alpha_{Is}|^2$ & \begin{tabular}{@{}l@{}}The sum of the square modulus of the spatial correlation \\ coefficient between each interference and the SOI\end{tabular} 	\tabularnewline
			\hline					
			$\alpha_I^2$ & A weighted sum of $|\alpha_{ps}|^2$ defined in \eqref{eq:alphaI} \tabularnewline
			\hline
			$\beta_p$ & \begin{tabular}{@{}l@{}}The phase difference between the $p$th interference and \\ the coefficient $\alpha_{ps}$ \end{tabular} \tabularnewline
			\hline		
			$\alpha_{w}$ & A weighted sum of $|\alpha_{ps}|^2$ defined in \eqref{eq:alphaPhi} \tabularnewline
			\hline
			$\Delta$ & A phase offset defined in \eqref{Phi_alpha} \tabularnewline	
			\hline
		\end{tabular}
\end{table}

From \eqref{eq:covx} and using \textit{Assumption 2}, we have
\begin{equation}\label{eq:Rinv1}
    {\textbf R}^{-1}=\frac{1}{\eta}\left(-\sum_{p=1}^{P}\frac{\pi_p}{\varepsilon_p\eta+\eta}{\textbf j}_p {\textbf j}_p^H+{\textbf I}_N\right),
\end{equation}
based on which, by further considering \textit{Assumption 1}, we obtain
\begin{equation}\label{eq:Rinv2}
    {\textbf R}^{-1}=-\frac{\pi}{\varepsilon\eta^2+\eta^2}\sum_{p=1}^{P}{\textbf j}_p {\textbf j}_p^H+\frac{1}{\eta}{\textbf I}_N.
\end{equation}
Upon substituting \eqref{eq:Rinv2} into \eqref{eq:SINRCapon1},  the output SINR of the Capon MVDR beamformer $\text{SINR}_{\scriptscriptstyle {\text {Capon}}}$ becomes
\begin{equation}
\text{SINR}_{\scriptscriptstyle {\text {Capon}}}
=-\frac{\pi\pi_s}{\varepsilon\eta^2+\eta^2}\sum_{p=1}^{P}{\textbf s}^H{\textbf j}_p {\textbf j}_p^H{\textbf s}+\frac{\pi_s}{\eta}{\textbf s}^H{\textbf s}.\label{eq:SINRCapon0}
\end{equation}
Upon inserting \eqref{power_s}, \eqref{eq:power_p} and \eqref{eq:spatial} into \eqref{eq:SINRCapon0}, the $\text{SINR}_{\scriptscriptstyle {\text {Capon}}}$ can be further simplified as
\begin{equation}
\text{SINR}_{\scriptscriptstyle {\text {Capon}}}=\varepsilon_s\left(1-\frac{\varepsilon}{\varepsilon+1}|\alpha_{Is}|^2\right),\label{eq:SINRCapon}
\end{equation}
%
where
\begin{align}
|\alpha_{Is}|^2 \triangleq \sum_{p=1}^{P}|\alpha_{ps}|^2 \label{eq:square modulus}
\end{align}
denotes the sum of the square modulus of the spatial correlation coefficient between each interference and the SOI, such that $0\leq|\alpha_{Is}|^2 \leq 1$ \cite{Lin1982}.

Now, similar to the derivation from \eqref{eq:Rinv1} to \eqref{eq:SINRCapon}, after some algebraic manipulations, the SINR for the WL MVDR beamformer defined in \eqref{eq:SINRdef} can be expressed by \eqref{eq:SINR!}, shown at the bottom of the page, where $\alpha_I^2$ is defined as
%
\newcounter{TempEqCnt}
\setcounter{TempEqCnt}{\value{equation}}
\begin{figure*}[hb]
\hrulefill
\setcounter{equation}{47}
\begin{align} \label{eq:SINR!}
	\text{SINR}_{\scriptscriptstyle {\text {MVDR}}}= \varepsilon_s \frac{\left[1+\varepsilon (2-|\alpha_{Is}|^2)+ \varepsilon^2 (1-|\gamma|^2)(1-|\alpha_{Is}|^2)\right]^2-\varepsilon^2|\alpha_I|^4 |\gamma|^2}{\big[(1+\varepsilon)^2-\varepsilon^2 |\gamma|^2\big]\big[1+\varepsilon (2-|\alpha_{Is}|^2)+ \varepsilon^2 (1-|\gamma|^2)(1-|\alpha_{Is}|^2)\big]}
	\end{align}
\hrulefill
\setcounter{equation}{49}
\begin{align} \label{eq:G!}
G = 1+\frac{|\gamma|^2 \varepsilon^2 \big[(1-|\alpha_{Is}|^2)|\alpha_{Is}|^2 \big((1+\varepsilon)^2-\varepsilon^2 |\gamma|^2\big) + (|\alpha_{Is}|^4-|\alpha_I|^4)(1+\varepsilon)\big]}{\big[1+\varepsilon(1-|\alpha_{Is}|^2)\big]\big[(1+\varepsilon)^2-\varepsilon^2 |\gamma|^2\big]\big[1+\varepsilon (2-|\alpha_{Is}|^2)+ \varepsilon^2 (1-|\gamma|^2)(1-|\alpha_{Is}|^2)\big]}
\end{align}
\setcounter{equation}{52}
\hrulefill
	\begin{align} \label{eq:lambdaI!}
\lambda_I=\frac{1+\varepsilon(2-|\alpha_{Is}|^2+|\gamma|\alpha_w)+\varepsilon^2(1-|\alpha_{Is}|^2)}{1+\varepsilon (2-|\alpha_{Is}|^2 + |\gamma|\alpha_{w})+ \varepsilon^2 (1-|\gamma|^2) (1-|\alpha_{Is}|^2)} \cdot \frac{1+\varepsilon (2-|\alpha_{Is}|^2)+ \varepsilon^2 (1-|\gamma|^2) (1-|\alpha_{Is}|^2)}{1+\varepsilon(2-|\alpha_{Is}|^2)+\varepsilon^2(1-|\alpha_{Is}|^2)}
	\end{align}
	\hrulefill
	\begin{align} \label{eq:lambdaQ!}
\lambda_Q=\frac{1+\varepsilon(2-|\alpha_{Is}|^2-|\gamma|\alpha_w)+\varepsilon^2(1-|\alpha_{Is}|^2)}{1+\varepsilon (2-|\alpha_{Is}|^2 - |\gamma|\alpha_{w})+ \varepsilon^2 (1-|\gamma|^2) (1-|\alpha_{Is}|^2)} \cdot \frac{1+\varepsilon (2-|\alpha_{Is}|^2)+ \varepsilon^2 (1-|\gamma|^2) (1-|\alpha_{Is}|^2)}{1+\varepsilon(2-|\alpha_{Is}|^2)+\varepsilon^2(1-|\alpha_{Is}|^2)}
	\end{align}
\end{figure*}
\begin{align}
\setcounter{equation}{48}
\alpha_I^2 \triangleq \sum_{p=1}^{P}|\alpha_{ps}|^2 e^{\jmath 2\beta_p},\label{eq:alphaI}
\end{align}
and $\beta_p \triangleq \phi_p-\phi_{ps}$ is the phase difference between the $p$th interference and the coefficient $\alpha_{ps}$. Thus, based on  \eqref{eq:SINRCapon} and \eqref{eq:SINR!}, the SINR gain $G$ defined in \eqref{eq:G_def} can be derived as \eqref{eq:G!}, shown at the bottom of the page.

In a similar way, after applying the \textit{Assumptions 1 and 2}, the SO noncircularity coefficient of the output IN of the Capon beamformer, $\gamma_{q,{\scriptscriptstyle {\text {Capon}}}}$, and that of the WL MVDR beamformer, $\gamma_{q,{\scriptscriptstyle {\text {MVDR}}}}$, can be respectively obtained as
\setcounter{equation}{50}
\begin{equation}\label{eq:gamma_qCapon}
\gamma_{q,{\scriptscriptstyle {\text {Capon}}}} = \frac{\gamma \varepsilon \alpha_I^2}{1+(2-\left|\alpha_{Is}\right|^2)\varepsilon+(1-\left|\alpha_{Is}\right|^2)\varepsilon^2},
\end{equation}	
\begin{equation}\label{eq:gamma_qMVDR}
\gamma_{q,{\scriptscriptstyle {\text {MVDR}}}} = \frac{\gamma \varepsilon \alpha_I^2}{1+(2-\left|\alpha_{Is}\right|^2)\varepsilon+(1-\left|\alpha_{Is}\right|^2)(1-\left|\gamma\right|^2)\varepsilon^2}.
\end{equation}
By substituting \eqref{eq:gamma_qCapon} and \eqref{eq:gamma_qMVDR} into \eqref{eq:dis_co}, we obtain explicit expressions for the distribution coefficients $\lambda_I$ and $\lambda_Q$, shown in \eqref{eq:lambdaI!} and \eqref{eq:lambdaQ!} at the bottom of the page, where
\setcounter{equation}{54}
\begin{equation}
\alpha_{w} \triangleq \sum_{p=1}^{P} |\alpha_{ps}|^2 \text{cos}(\delta +2\beta_p),\label{eq:alphaPhi}
\end{equation}
is a weighted sum of $|\alpha_{ps}|^2$ and it is straightforward that $0<\left|\alpha_{w}\right|\leq|\alpha_{Is}|^2$.

Now, equations \eqref{eq:SINR!} to \eqref{eq:alphaPhi} pave the way to understand the statistical behavior of both the beamformers in the I and Q channels. Without loss of generality, we next consider the following three scenarios, covering all the spatial relationships between the SOI and the interferences.
\subsubsection{The SOI is orthogonal to all of the interferences}
In this special case, according to \textit{Assumption 2}, the spatial correlation coefficient, $\alpha_{ps}$ in \eqref{eq:spatial}, becomes $\alpha_{ps}=0$ for all the interferences. From \eqref{eq:square modulus}, \eqref{eq:alphaI} and \eqref{eq:alphaPhi}, this gives
\begin{equation}\label{eq:orthCondition}
\alpha_I=\alpha_{Is}=\alpha_{w}=0,
\end{equation}
and consequently, from \eqref{eq:SINRCapon}, \eqref{eq:SINR!}, and \eqref{eq:G!} to \eqref{eq:lambdaQ!}, we have
\begin{equation}\label{eq:resultorthogonal}
\begin{gathered}
\text{SINR}_{\scriptscriptstyle {\text {Capon}}} = \text{SINR}_{\scriptscriptstyle {\text {MVDR}}} = \varepsilon_s, \\
\gamma_{q,{\scriptscriptstyle {\text {Capon}}}}=\gamma_{q,{\scriptscriptstyle {\text {MVDR}}}}=0,\\
G=G_I = G_Q = 1.
\end{gathered}
\end{equation}

\textit{Remark 2:} Equation \eqref{eq:resultorthogonal} indicates that both the beamformers yield an identical overall SINR performance, which is also equivalent in the respective I and Q channels, at the power ratio between the SOI and the noise, $\varepsilon_s$. This is as expected, because in this case, the output INs of both the beamformers, $q_{\scriptscriptstyle{\text {MVDR}}}(t)$ and $q_{\scriptscriptstyle{\text {Capon}}}(t)$, becomes SO circular, as evidenced by $\gamma_{q,{\scriptscriptstyle {\text {Capon}}}}=\gamma_{q,{\scriptscriptstyle {\text {MVDR}}}}=0$.
\subsubsection{The SOI is a linear combination of the interferences}
Based on \eqref{eq:spatial}, we observe that the spatial correlation coefficient, ${\alpha_{ps}}$, is essentially the normalized projection of the steering vector of  the $p$th interference, ${\textbf j}_p$, on the steering vector of the SOI, ${\textbf s}$. In this sense, ${\textbf s}$ can be represented as a linear combination of interference vectors, ${\textbf j}_p, p \in \{1,\ldots,P\}$. By substituting \eqref{eq:spatial} into \eqref{eq:square modulus} and employing the orthogonality condition in \textit{Assumption 2}, we obtain
%
\begin{equation}\label{eq:squaremodulusequalsone}
|\alpha_{Is}|=1.
\end{equation}
%
Now, a substitution of \eqref{eq:squaremodulusequalsone} into \eqref{eq:SINRCapon}, \eqref{eq:SINR!}, and \eqref{eq:G!} to \eqref{eq:lambdaQ!} yields
\begin{equation}\label{eq:col_dis}
\begin{gathered}
\text{SINR}_{\scriptscriptstyle {\text {Capon}}}= \frac{\varepsilon_s}{\varepsilon+1},\\
\text{SINR}_{\scriptscriptstyle {\text {MVDR}}} = \varepsilon_s \frac{(1+\varepsilon)^2-\varepsilon^2|\alpha_I|^4|\gamma|^2}{[(1+\varepsilon)^2-\varepsilon^2|\gamma|^2](1+\varepsilon)},\\
\gamma_{q,{\scriptscriptstyle {\text {Capon}}}}=\gamma_{q,{\scriptscriptstyle {\text {MVDR}}}}=\gamma \varepsilon \alpha_I^2,\\
G=G_I=G_Q=|\alpha_I|^4\!\!+\!\!\frac{(1\!-\!|\alpha_I|^4)(1+\varepsilon)^2}{(1\!+\!\varepsilon)^2\!-\!\varepsilon^2|\gamma|^2}.\\
\end{gathered}
\end{equation}

\textit{Remark 3:} In this case, the MVDR beamformer outperforms the Capon one, since its overall performance gain, $G$, monotonically increases with the SO noncircularity rate of interferences for $0 < |\gamma| < 1$, and hence, $G > 1$. Moreover, the output INs of both beamformers, $q_{\scriptscriptstyle{\text {MVDR}}}(t)$ and $q_{\scriptscriptstyle{\text {Capon}}}(t)$, become SO noncircular but still with the same SO noncircularity coefficient, $\gamma \varepsilon \alpha_I^2$. Consequently, the distribution coefficients in \eqref{eq:lambdaI!} and \eqref{eq:lambdaQ!} become $\lambda_I = \lambda_Q = 1$, which in turn makes the performance advantage of the WL MVDR beamformer over the Capon one equally carried in the I and Q channels, and these are still equivalent to the overall gain, i.e, $G_I=G_Q=G$.
%

%
\subsubsection{The most general case}
When the SOI is neither orthogonal to the interferences nor linearly combined by them, according to \eqref{eq:square modulus}, we have $0< |\alpha_{Is}| < 1$. Therefore, based on \eqref{eq:gamma_qCapon} and \eqref{eq:gamma_qMVDR}, the output INs, $q_{\scriptscriptstyle{\text {MVDR}}}(t)$ and $q_{\scriptscriptstyle{\text {Capon}}}(t)$, are both SO noncircular, but with different SO noncircularity coefficients.

Suppose that all the interferences are much stronger than the noise, i.e., $\varepsilon \gg 1$ \cite{Chevalier2007}, which makes the terms in \eqref{eq:SINR!} $\varepsilon$ and $\varepsilon^2$ negligible as compared with other high-order terms of $\varepsilon$, so that
%
%
%
\begin{equation}
\text{SINR}_{\scriptscriptstyle {\text {MVDR}}} \!\approx\! \varepsilon_s \left[1\!-\!|\alpha_{Is}|^2\!+\!\frac{|\alpha_{Is}|^2(\varepsilon\!+\!1)}{(1\!-\!|\gamma|^2)\varepsilon^2\!+\!2\varepsilon\!+\!1}\right].\!\label{eq:SINR_appro}
\end{equation}
%
Upon dividing \eqref{eq:SINR_appro} by \eqref{eq:SINRCapon} we have
%
\begin{equation}
G \!\approx\! 1\!+\!\frac{|\alpha_{Is}|^2|\gamma|^2\varepsilon}{(1\!-\!|\alpha_{Is}|^2)(1\!-\!|\gamma|^2)\varepsilon^2\!+\!(2\!-\!|\alpha_{Is}|^2)\varepsilon\!+\!1}.\label{eq:G_appro}
\end{equation}
%

The result in \eqref{eq:G_appro} extends the analysis in \cite{Chevalier2007}, and shows that for the reception of the unknown SOI in the presence of an arbitrary number of orthogonal interferences, the SINR gain $G$ is a monotonically increasing function of the SO noncircularity rate $|\gamma|$ of the interferences, so that, $G>1$ for $0 < |\gamma| < 1$.

In a similar way, the individual distribution coefficients, $\lambda_I$ in \eqref{eq:lambdaI!} and $\lambda_Q$  in \eqref{eq:lambdaQ!}, can be approximated as
%
%
%
\begin{align}
\lambda_I &\approx 1-\frac{|\gamma|^3\alpha_{w}}{\varepsilon(1-|\alpha_{Is}|^2)(1-|\gamma|^2)},\label{eq:lambdaI_generic_approx}\\
\lambda_Q &\approx 1+\frac{|\gamma|^3\alpha_{w}}{\varepsilon(1-|\alpha_{Is}|^2)(1-|\gamma|^2)},\label{eq:lambdaQ_generic_approx}
\end{align}
%
which are both monotonic functions of $|\gamma|$, although with opposing monotonicities.

By multiplying \eqref{eq:lambdaI_generic_approx} and \eqref{eq:lambdaQ_generic_approx} with \eqref{eq:G_appro} respectively, we obtain
\begin{align}
G_I &\approx 1+\frac{-\alpha_{w}|\gamma|^3+|\alpha_{Is}|^2|\gamma|^2}{\varepsilon(1-|\alpha_{Is}|^2)(1-|\gamma|^2)},\label{eq:GI_generic_approx}\\
%
%
G_Q &\approx 1+\frac{\alpha_{w}|\gamma|^3+|\alpha_{Is}|^2|\gamma|^2}{\varepsilon(1-|\alpha_{Is}|^2)(1-|\gamma|^2)}.\label{eq:GQ_generic_approx}
\end{align}

\textit{Remark 4:} Owing to the fact that $0<\left|\alpha_{w}\right|\leq|\alpha_{Is}|^2$, it is guaranteed, from \eqref{eq:GI_generic_approx} and \eqref{eq:GQ_generic_approx}, that $G_I > 1$ and $G_Q>1$, regardless of the values of $\alpha_{w}$ and $\gamma$. Therefore, in the most general case, the WL MVDR beamformer always provides SINR gains over the Capon one in both the I and Q channels. Moreover, as proved in Appendix \ref{App:AppendixA}, $G_I$ and $G_Q$ are both monotonically increasing functions of $|\gamma|$, although their increasing slopes are different, dependent on the sign of the coefficient $\alpha_{w}$.
%


Through the above analysis, the links between the SO noncircularity rate $|\gamma|$ of the interferences and the SINR gains in the I and Q channels have been discussed for the most general case. Moreover, the coefficient $\alpha_{w}$ is shown to play a pivotal role in determining the increasing slope of $G_I$ and $G_Q$ w.r.t. $|\gamma|$. As defined in \eqref{eq:alphaPhi}, the value of $\alpha_{w}$ is determined by the SO noncircularity phase $\delta$ of the interferences, and it can be further expanded as
\begin{align} \label{Phi_alpha}
&\alpha_{w} \!\!=\!\! \sum_{p=1}^{P} |\alpha_{ps}|^2 \text{cos}(\delta + 2\beta_p) \nonumber\\
&\!\!=\!\! \left(\sum_{p=1}^{P} |\alpha_{ps}|^2 \text{cos}2\beta_p\right)\text{cos}\delta-\left(\sum_{p=1}^{P} |\alpha_{ps}|^2 \text{sin}2\beta_p\right)\text{sin}\delta \nonumber\\
&\!\!=\!\! \sqrt{\left(\sum_{p=1}^{P} |\alpha_{ps}|^2 \text{cos}2\beta_p\right)^2+\left(\sum_{p=1}^{P} |\alpha_{ps}|^2 \text{sin}2\beta_p\right)^2}\cdot\text{cos}(\delta\!+\!\Delta),
\end{align}
where $\Delta$ satisfies
\begin{align}
\text{cos}\Delta = \frac{\sum_{p=1}^{P} |\alpha_{ps}|^2 \text{cos}2\beta_p}{\sqrt{\left(\sum_{p=1}^{P} |\alpha_{ps}|^2 \text{cos}2\beta_p\right)^2+\left(\sum_{p=1}^{P} |\alpha_{ps}|^2 \text{sin}2\beta_p\right)^2}}, \nonumber
\\ \text{sin}\Delta = \frac{\sum_{p=1}^{P} |\alpha_{ps}|^2 \text{sin}2\beta_p}{\sqrt{\left(\sum_{p=1}^{P} |\alpha_{ps}|^2 \text{cos}2\beta_p\right)^2+\left(\sum_{p=1}^{P} |\alpha_{ps}|^2 \text{sin}2\beta_p\right)^2}}.\nonumber
\end{align}
%

\textit{Remark 5:} The coefficient $\alpha_{w}$ is a cosine function of the SO noncircularity phase $\delta$ of the interferences. Therefore, based on \eqref{eq:GI_generic_approx} and \eqref{eq:GQ_generic_approx}, both $G_I$ and $G_Q$ vary cyclically over $\delta$. Table \ref{tab:Phi} summarizes the effect of $\delta$ on $G_I$ and $G_Q$ when $\delta$ varies within each period $(-\Delta+2k\pi, -\Delta+2\pi+2k\pi)$, where $k$ is an arbitrary integer.
\begin{table}[t!]
	\centering
	\caption{Effect of the SO noncircularity phase $\delta$ of the interferences on the SINR gains in the I and Q channels, $G_I$ and $G_Q$.}
	\label{tab:Phi}
	
	\begin{tabular}{|@{\hskip1pt}c@{\hskip1pt}|@{\hskip2pt}l@{\hskip2pt}|@{\hskip2pt}l@{\hskip2pt}|}
		\hline
		\thead{$\delta$} & \thead{$G_I$} &\thead{$G_Q$}\\
		\hline
		($-\Delta\!\!-\!\!\frac{\pi}{2}\!\!+\!\!2k\pi$,$-\Delta\!\!+\!\!2k\pi$) & \begin{tabular}[c]{@{}l@{}}Decrease on $\delta$\\Increase flatly on $\left|\gamma\right|$ \end{tabular} & \begin{tabular}[c]{@{}l@{}}Increase on $\delta$\\Increase fastly on $\left|\gamma\right|$ \end{tabular}  \\
		\hline
		($-\Delta\!\!+\!\!2k\pi$,$-\Delta\!\!+\!\!\frac{\pi}{2}\!\!+\!\!2k\pi$) & \begin{tabular}[c]{@{}l@{}}Increase on $\delta$\\Increase flatly on $\left|\gamma\right|$  \end{tabular}   & \begin{tabular}[c]{@{}l@{}}Decrease on $\delta$\\Increase fastly on $\left|\gamma\right|$  \end{tabular}  \\
		\hline
		($-\Delta\!\!+\!\!\frac{\pi}{2}\!\!+\!\!2k\pi$,$-\Delta\!\!+\!\!\pi\!\!+\!\!2k\pi$) & \begin{tabular}[c]{@{}l@{}}Increase on $\delta$\\Increase fastly on $\left|\gamma\right|$  \end{tabular}   & \begin{tabular}[c]{@{}l@{}}Decrease on $\delta$\\Increase flatly on $\left|\gamma\right|$  \end{tabular}  \\
		\hline
		($-\Delta\!\!+\!\!\pi\!\!+\!\!2k\pi$,$-\Delta\!\!+\!\!\frac{3\pi}{2}\!\!+\!\!2k\pi$) & \begin{tabular}[c]{@{}l@{}}Decrease on $\delta$\\Increase fastly on $\left|\gamma\right|$  \end{tabular}  & \begin{tabular}[c]{@{}l@{}}Increase on $\delta$ \\Increase flatly on $\left|\gamma\right|$ \end{tabular}   \\
		\hline
	\end{tabular}
\end{table}
\begin{figure*}[t!]
	\centering
	\subfigure[]{
		\label{fig: GI_2I2N} 
		\begin{minipage}[]{0.5\textwidth}
			\centering
			\includegraphics[width=0.9\textwidth]{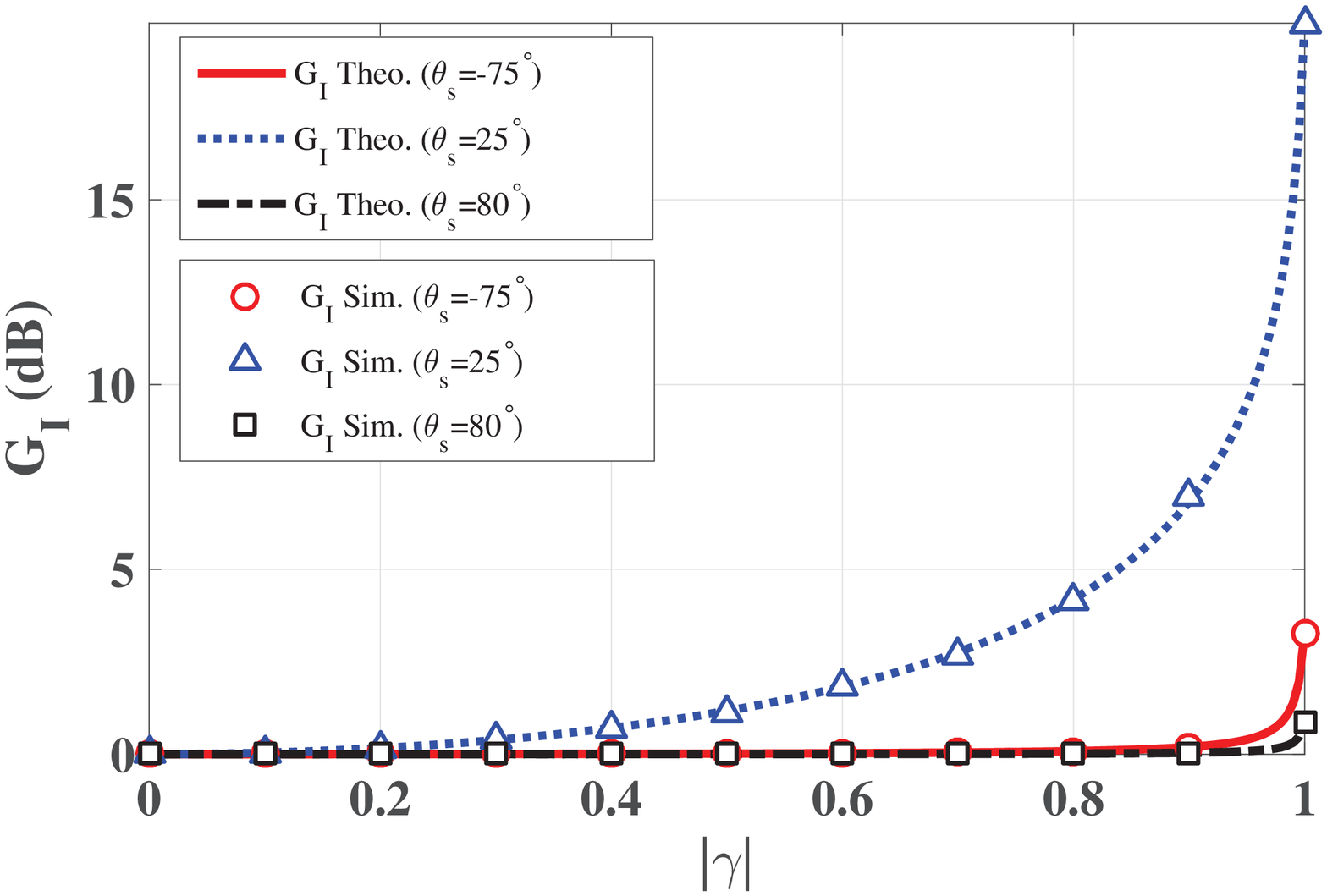}
		\end{minipage}
	}%
	\subfigure[]{
		\label{fig:GQ_3I6N} 
		\begin{minipage}[]{0.5\textwidth}
			\centering
			\includegraphics[width=0.9\textwidth]{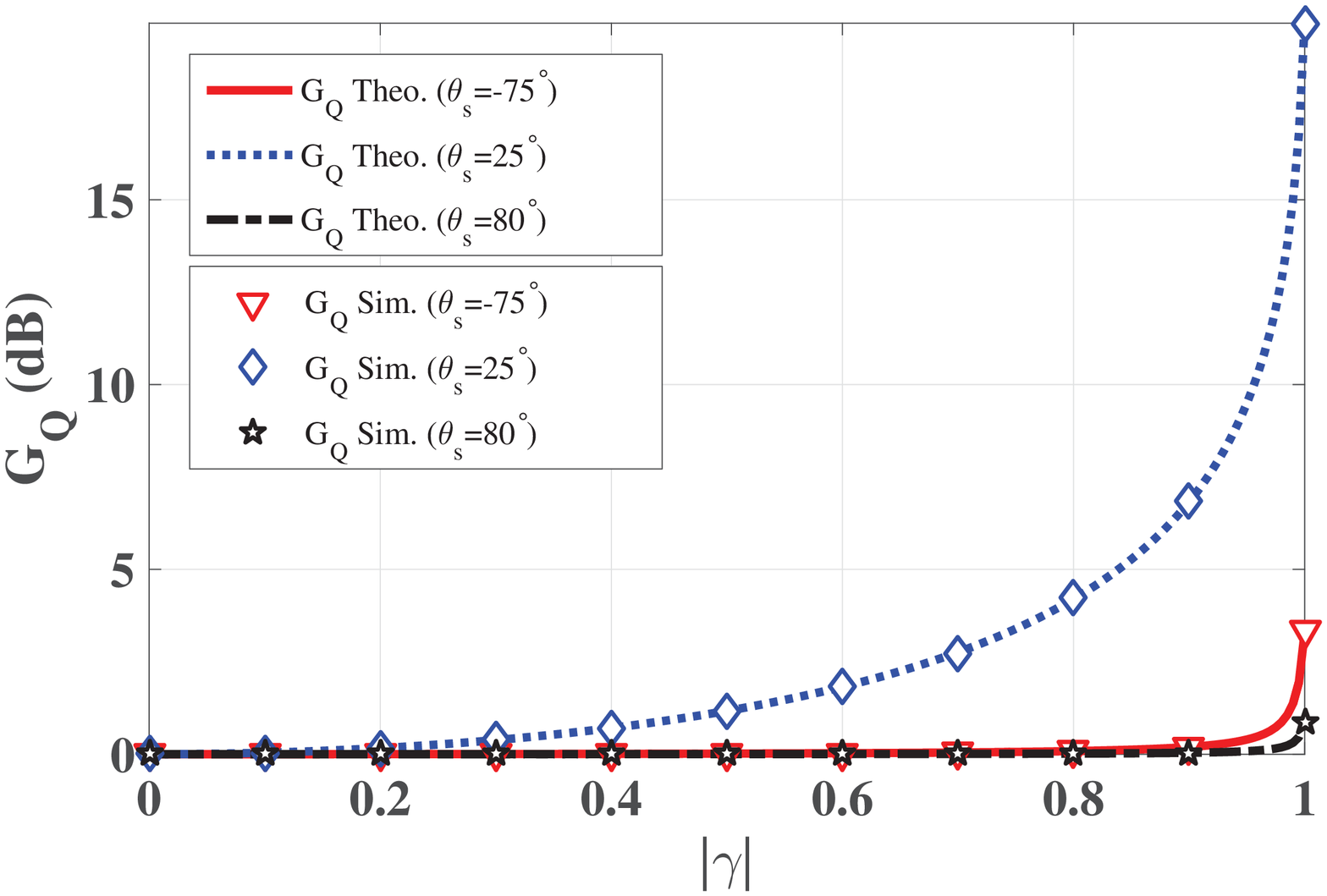}
		\end{minipage}
	}
	\caption{Theoretical and simulated individual SINR gains in the I and Q channels, $G_I$ and $G_Q$, as functions of the noncircularity rate $|\gamma|$ of the interferences, for several DOAs of the SOI $\theta_s$ ($-75^\circ$, $25^\circ$ and $80^\circ$). The beamformer parameters used for illustration were $N=2$, $P=2$, $\theta_1=0^\circ$, $\theta_2=90^\circ$. (a) $G_I$, and (b) $G_Q$.}
	\label{fig:GIGQ_2I2N} 
\end{figure*}
\begin{figure*}[t!]
	\centering
	\subfigure[]{
		\label{fig:GIGQ_3I6N} 
		\begin{minipage}[]{0.5\textwidth}
			\centering
			\includegraphics[width=0.9\textwidth]{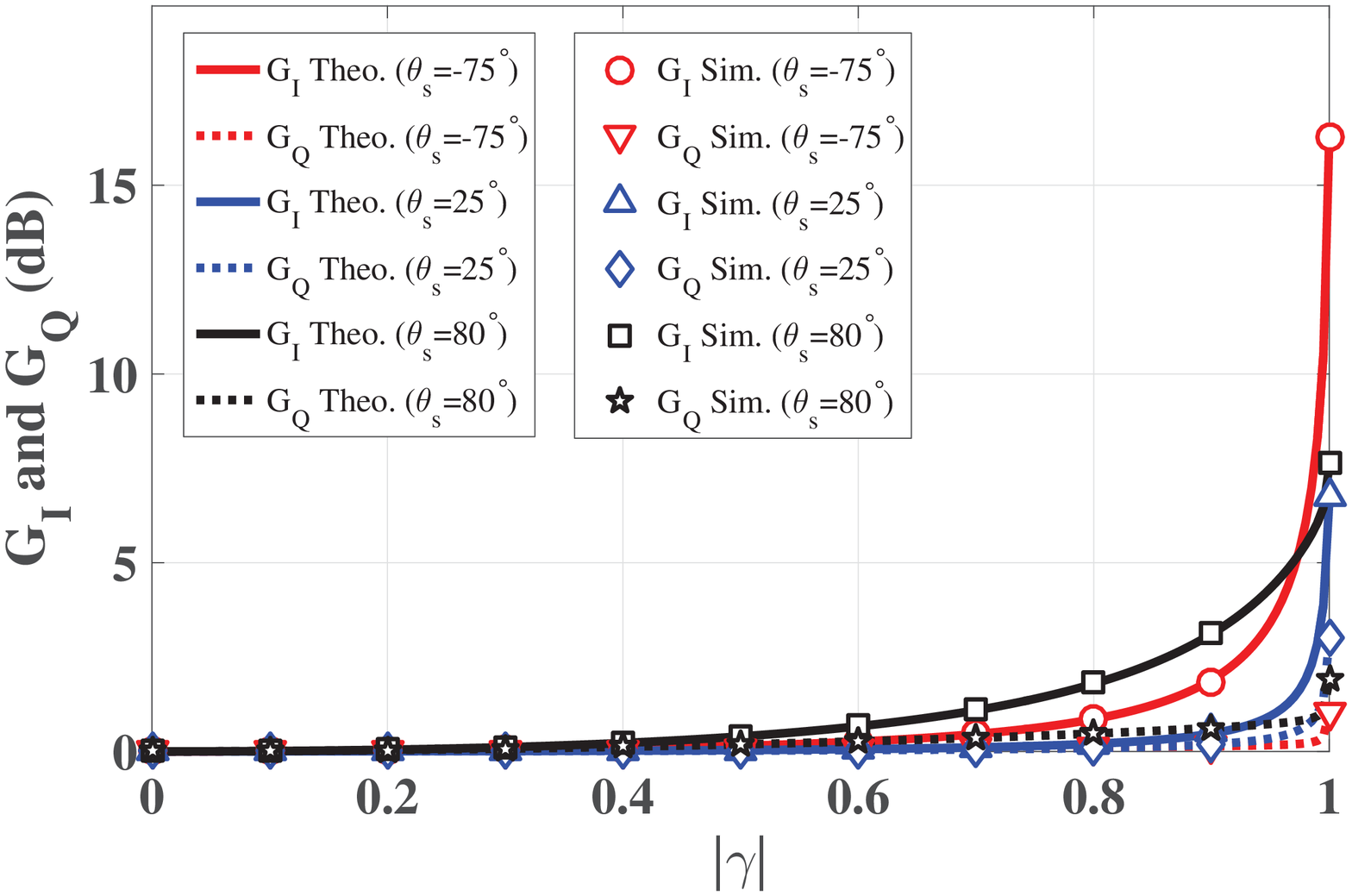}
		\end{minipage}
	}%
	\subfigure[]{
		\label{fig:GIGQ_3I6N2} 
		\begin{minipage}[]{0.5\textwidth}
			\centering
			\includegraphics[width=0.9\textwidth]{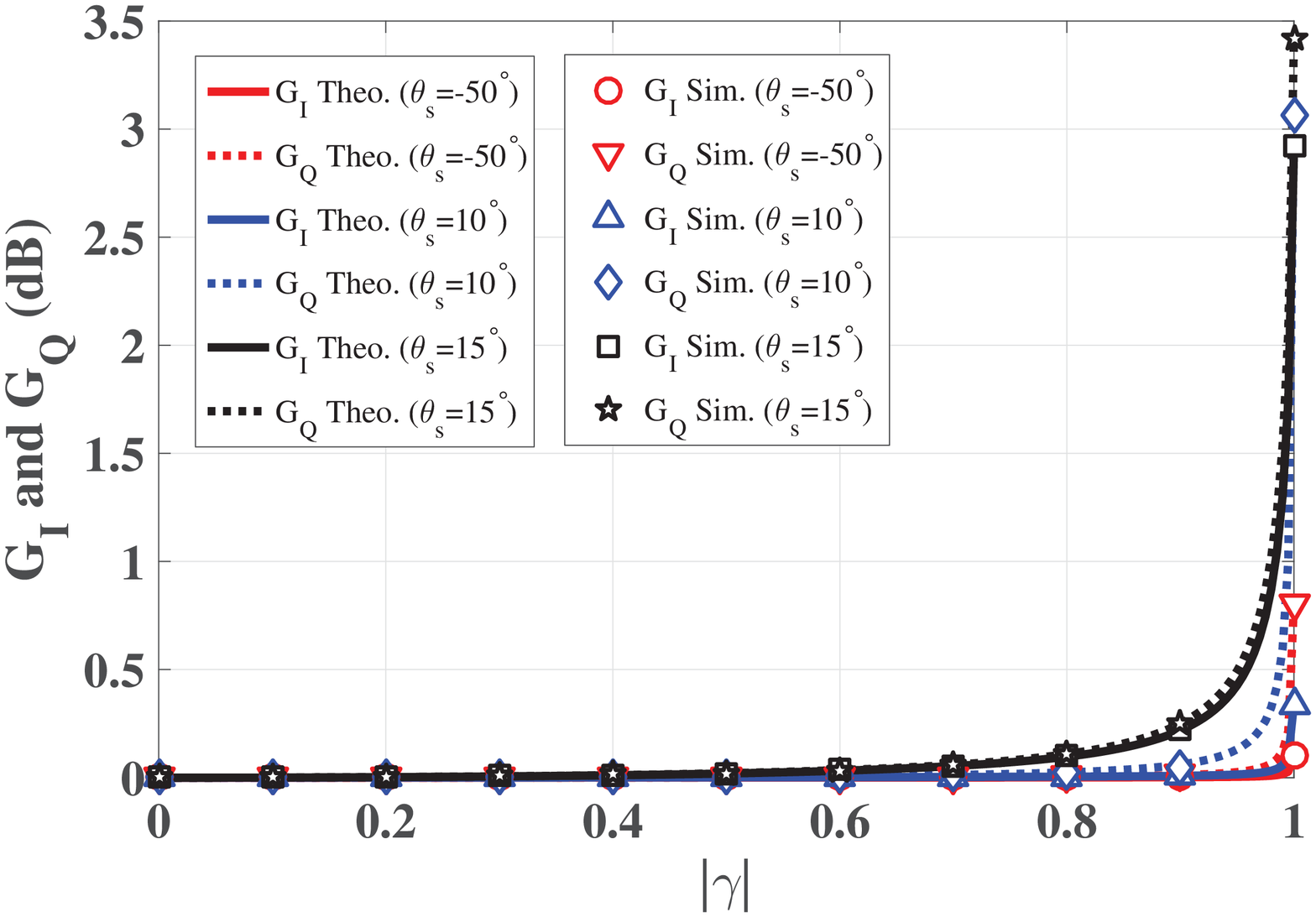}
		\end{minipage}
	}
	\caption{Theoretical and simulated individual SINR gains in the I and Q channels, $G_I$ in and $G_Q$, as functions of the SO noncircularity rate $|\gamma|$ of the interferences, where the beamformer parameters were $N=6$, $P=3$, $\theta_1=19^\circ$, $\theta_2=42^\circ$ and $\theta_3=90^\circ$. Two sets of DOAs of the SOI $\theta_s$ used for illustration were (a) ($-75^\circ$, $25^\circ$, $80^\circ$), and (b) ($-50^\circ$, $10^\circ$, $15^\circ$).}
	\label{fig:GIGQ_3I6Ncombo} 
\end{figure*}
\section{Numerical Simulations}
Simulations were conducted in the MATLAB programming environment to validate the proposed full SO performance analysis of both the WL MVDR and Capon beamformers. We first considered the DOA estimation of noncircular signals, compliant to the unbalanced quadrature phase shift keying (UQPSK) constellation. Then, a more practical beamforming scenario which considers radio frequency (RF) I/Q imbalanced QPSK signals was examined.
\subsection{DOA Estimation of UQPSK Signals}\label{sec:UQPSK}
In this set of simulations, the SOI, $s_c(t)$, and the $p$th interference, $m_{cp}(t)$, were both UQPSK modulated, given by \cite{Xu2013}
\begin{align}
&s_{c}(t) = \xi\left[\sum_n\rho_s(n)g(t-n/f_s)\cos(2\pi f_s t+\delta_s)\right]\nonumber\\
&+j(1-\xi)\left[\sum_n\sigma_s(n)g(t-n/f_s)\sin(2\pi f_s t+\delta_s)\right],\label{Eq:UQPSK_SOI}\\
%
%
&m_{cp}(t) = \xi\left[\sum_n\rho_{cp}(n)g(t-n/f_p)\cos(2\pi f_p t+\delta_p)\right]\nonumber\\
&+j(1-\xi)\left[\sum_n\sigma_{cp}(n)g(t-n/f_p)\sin(2\pi f_p t+\delta_p)\right],\label{Eq:UQPSK_intf}
\end{align}
where $\xi \in [0,1]$ is the equivalence factor which controls their SO noncircularity coefficients, the SO noncircularity phases, $\delta_s$ and $\delta_p$, represent respectively the residual Doppler shift (frequency offset) of the SOI and the $p$th interference at the output of the demodulator \cite{Clark1989}, and $g(t)$ is a rectangular pulse-shaped filter, given by
\begin{equation}
g(t)=\left\{ {\begin{array}{*{20}{c}}
	{1\quad 0<t<1/f_s},\\
	{0\;\quad\quad {\rm elsewhere}.}
	\end{array}} \right.
\end{equation}
The variables $\rho_s(n)$ and $\sigma_s(n)$ in \eqref{Eq:UQPSK_SOI} are respectively the $n$th in-phase and quadrature components of the SOI, while $\rho_{cp}(n)$ and $\sigma_{cp}(n)$ in \eqref{Eq:UQPSK_intf} are respectively the $n$th in-phase and quadrature components of the $p$th interference. All these were independent and identically distributed random variables, subject to a two-point distribution with an equal probability for $1$ and $-1$. Throughout the simulations, the signal-to-noise ratio $\text{SNR} \triangleq {\pi_s}/{\eta}$ was set to 10 dB, while the interference-to-noise ratio $\text{INR} \triangleq {\pi}/{\eta}$ was set to 20 dB \cite{Chevalier2007}.
%
%
\begin{figure}[t!]
	\centering
	\centering
	\includegraphics[width=0.45\textwidth]{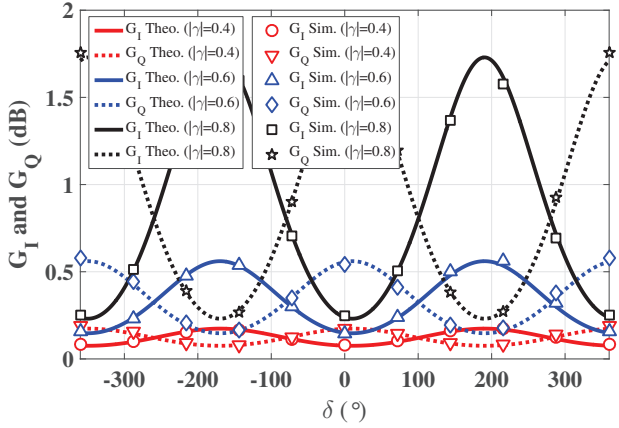}
	\caption{Theoretical and simulated individual SINR gains in the I and Q channels, $G_I$ and $G_Q$, as functions of the SO noncircularity phase $\delta$ of the interferences, for $|\gamma|$ = 0.4, 0.6, and 0.8. The system parameters of beamformers were: $\theta_s=85^\circ$, $N=16$, $P=4$, $\theta_1=14^\circ$, $\theta_2=30^\circ$, $\theta_3=49^\circ$ and $\theta_4=90^\circ$.}
	\label{fig:SINR_Phi_2I2N}
\end{figure}
\begin{figure}[t!]
	\centering
	\subfigure[]{
		\label{fig:RFA_WL_MVDR:Beamformer} 
		\begin{minipage}[]{0.5\textwidth}
			\centering
			\includegraphics[width=0.9\textwidth]{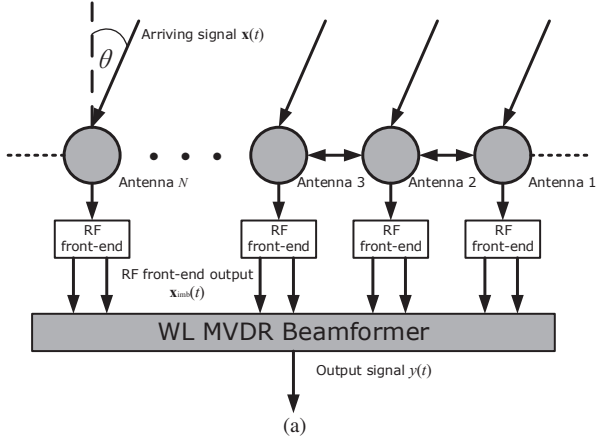}
		\end{minipage}
	}%
	\\
	\subfigure[]{
		\label{fig:RFA_WL_MVDR:IQI} 
		\begin{minipage}[]{0.5\textwidth}
			\centering
			\includegraphics[width=0.9\textwidth]{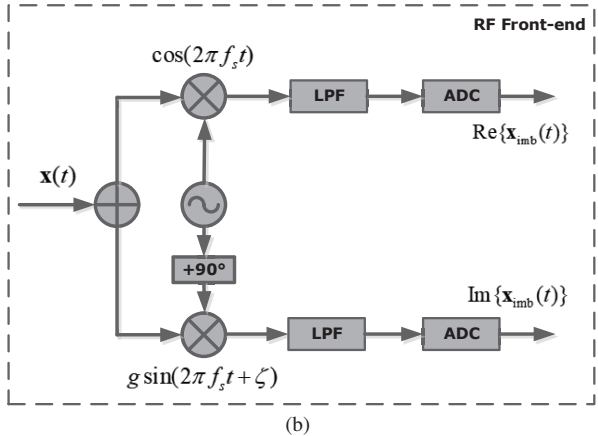}
		\end{minipage}
	}
	\caption{The WL MVDR beamformer with a uniform linear array implementation. The flow of the arriving signal is presented by arrows. (a) The beamforming structure, and (b) The I/Q imbalanced RF front-end.}
	\label{fig:RFA_WL_MVDR} 
\end{figure}
\begin{figure}[t!]
	\centering
	\includegraphics[width=0.45\textwidth]{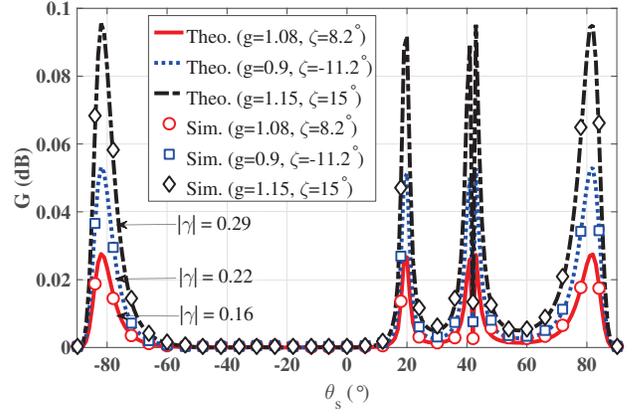}
	\caption{The SINR gain of the WL MVDR beamformer as a function of the DOA of the SOI $\theta_s$ for three types of I/Q imbalanced antenna arrays.}
	\label{fig:IQGain} 
\end{figure}

We first considered two different spatial relationships between the SOI and the interferences. In the first case, a uniform linear array with $N=2$ omnidirectional sensors spaced half a wavelength apart was used and the desired signal was corrupted by $P=2$ interferences, whose DOAs were equal to $\theta_1=0^\circ$ and $\theta_2=90^\circ$, respectively. In this way, since $P=N$, the steering vector of the SOI, ${\textbf s}$, can be represented as a linear combination of the interference vectors $\{{\textbf j}_1,{\textbf j}_2\}$. The second set of beamformer parameters fit the most general case, where the number of sensors and interferences were respectively set to $N=6$ and $P=3$, and the DOAs of the three interferences were equal to $\theta_1=19^\circ$, $\theta_2=42^\circ$ and $\theta_3=90^\circ$, respectively. In both cases, the theoretical SINR gains in the I and Q channels of the WL beamformer over the Capon one, that is, $G_I$ and $G_Q$, were obtained by multiplying $G$ in \eqref{eq:G!} with $\lambda_I$ in \eqref{eq:lambdaI!} and $\lambda_Q$ in \eqref{eq:lambdaQ!} respectively, while the simulated SINR gains were obtained by using up to 20,000 UQPSK modulated signals. Both the theoretical and simulated $G_I$ and $G_Q$ are plotted in Fig. \ref{fig:GIGQ_2I2N} and Fig. \ref{fig:GIGQ_3I6Ncombo} as functions of the SO noncircularity rate $|\gamma|$ of the interferences. The set of DOA situations of the SOI $\theta_s$ for investigation was ($-75^\circ$, $-25^\circ$, $80^\circ$) in Fig. \ref{fig:GIGQ_2I2N} and Fig. \ref{fig:GIGQ_3I6N}, and was ($-50^\circ$, $10^\circ$, $15^\circ$) in Fig. \ref{fig:GIGQ_3I6N2}, while the SO noncircularity phase  of the interferences was fixed at $\delta_p =\delta=150^\circ$. Observe from Fig. \ref{fig:GIGQ_2I2N} and Fig. \ref{fig:GIGQ_3I6Ncombo} that the analytical SINR gains were closely matched by their empirical counterparts. Moreover, in all the cases, $G_I > 0$ dB and $G_Q > 0$ dB, and the performance advantages of the WL beamformer over the Capon one in both the channels were more pronounced as $|\gamma|$ increased. Besides, by comparing the results in Fig. \ref{fig:GIGQ_2I2N}(a) and Fig. \ref{fig:GIGQ_2I2N}(b), we found that when the SOI is a linear combination of the interferences, $G_I$ and $G_Q$ are identical, which is in line with the analysis in \textit{Remark 3}. However, this is not the case in Fig. \ref{fig:GIGQ_3I6N}, where $G_I$ is always larger than $G_Q$ for different DOAs of the SOI $\theta_s$. This phenomenon can be explained by \textit{Remark 4}, because in this specific experimental setting, the coefficient $\alpha_{w}<0$, so that the increasing speed of $G_I$ w.r.t. $|\gamma|$ is faster than that of $G_Q$. On the other hand, there also exists a situation where $G_Q$ increases faster than $G_I$ w.r.t. $|\gamma|$. This is supported by Fig. \ref{fig:GIGQ_3I6N2}, where all the considered DOA situations of the SOI $\theta_s$, ($-50^\circ,10^\circ, 15^\circ$), gave a positive $\alpha_{w}$. Both Fig. \ref{fig:GIGQ_3I6N} and Fig. \ref{fig:GIGQ_3I6N2}  have well validated our approximation on $G_I$ and $G_Q$ in the most general case.

We next fixed the DOA of the SOI $\theta_s$ at $85^\circ$, and set  $|\gamma|$ = 0.4, 0.6 and 0.8 respectively, in order to examine how $G_I$ and $G_Q$ adapted to the changes in  the SO noncircularity phase $\delta$ of the interferences. Again, very good agreement between the theoretical and simulated individual SINR gains can be observed in Fig. \ref{fig:SINR_Phi_2I2N}. Especially, for different values of $|\gamma|$, the individual SINR gains in both the I and Q channels, $G_I$ and $G_Q$, vary periodically over $\delta$. Moreover, as discussed in \textit{Remark 5} and Table \ref{tab:Phi}, given the same value of $\delta$, they always exhibit different increasing slopes against $|\gamma|$, in each quarter of the cycle $2\pi$, while it is always guaranteed that $G_I$ and $G_Q$ are above the unity (0 dB).

\subsection{DOA Estimation of QPSK Signals in Presence of RF I/Q Impairment}
Direct-conversion receivers have been widely adopted in massive communication systems owing to their simple structure and low cost. Unfortunately, they suffer from a common RF imperfection, called I/Q imbalance, due to non-ideal properties of RF mixers. The I/Q imbalance makes the arriving signal noncircular, even though typical communication waveforms, such as QPSK and quadrature amplitude modulation, are of circular nature \cite{Anttila2008,Li2017}. To automatically suppress the unwanted effects of the RF I/Q imbalance on the beamsteering, the WL MVDR beamformer was employed in \cite{Hakkarainen2013}, with the detailed implementation depicted in Fig. \ref{fig:RFA_WL_MVDR}. 
Suppose the carrier residue of the SOI is equivalent to that of the interferences, i.e., $f_s=f_p$, then, the input of the WL MVDR beamformer becomes \cite{Anttila2008}
\begin{equation}\label{Eq:ximbt}
	{\bf x}_{\scriptscriptstyle{\text {imb}}} (t) = \mu {\textbf x}(t) + \nu {\textbf x}^*(t),
\end{equation}
where the original arriving signal ${\textbf x}(t)$ is QPSK modulated, and  $\mu$ and $\nu$ are defined as
\begin{equation}
\mu\triangleq\frac{1+ge^{-\jmath\zeta }}{2}, \quad\quad
\nu\triangleq\frac{1-ge^{\jmath\zeta}}{2}.\label{eq:IQphaseamp}
\end{equation}
The variables $g$ and $\zeta$ in \eqref{eq:IQphaseamp} represent respectively the relative amplitude and phase imbalances in each RF front-end. Then, according to \eqref{Eq:ximbt}, by absorbing the I/Q imbalance influence into the original arriving signal ${\bf x}(t)$, the SO noncircularity coefficients of the SOI and the $p$th interference, that is, $\gamma_s$ and $\gamma_p$, can be equivalently expressed as
\begin{figure*}[t!]
	\centering
	\subfigure[]{
		\label{fig:IQGIGQ1} 
		\begin{minipage}[]{0.31\textwidth}
			\centering
			\includegraphics[width=\textwidth]{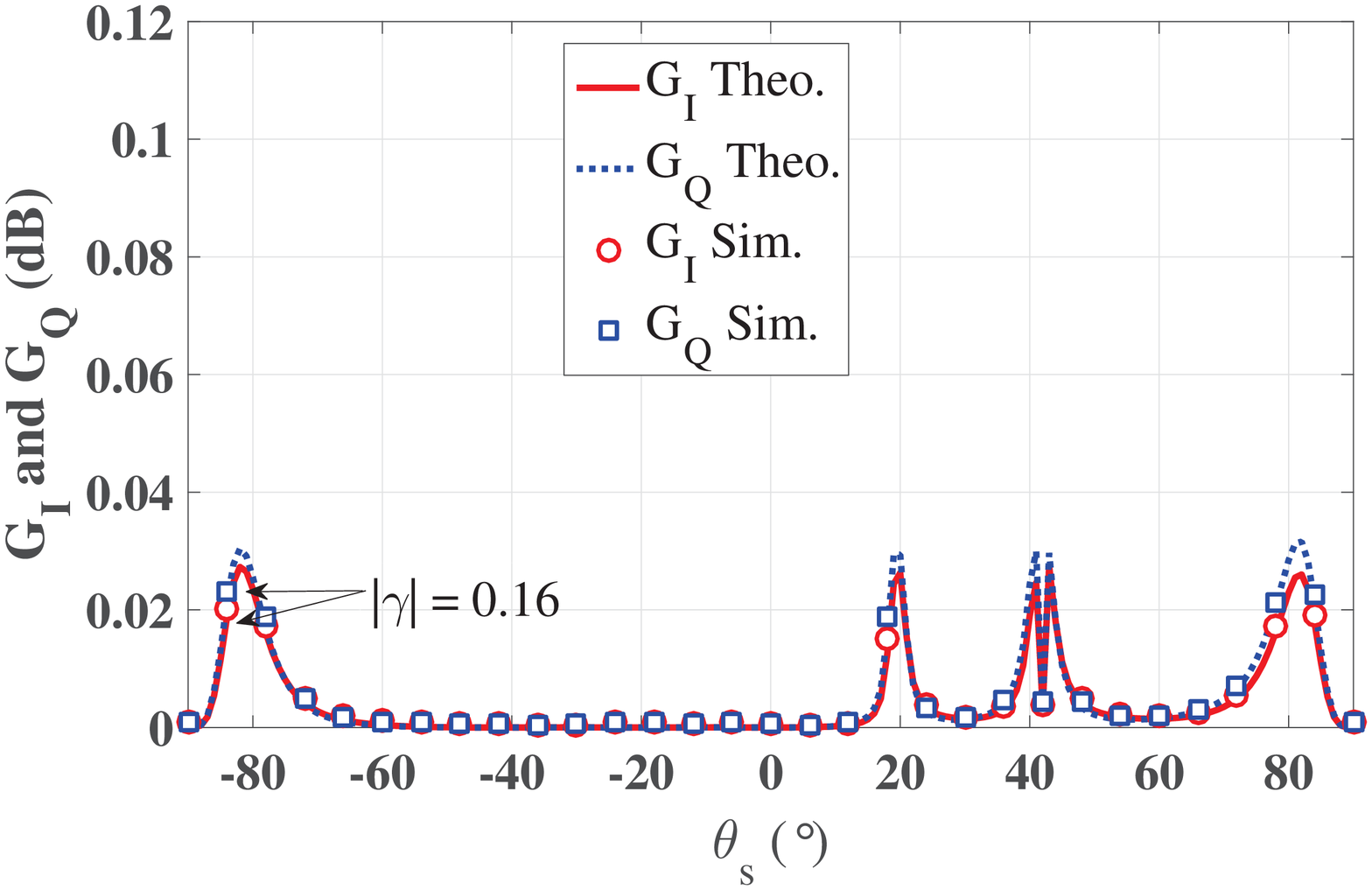}
		\end{minipage}
	}%
	\subfigure[]{
		\label{fig:IQGIGQ2} 
		\begin{minipage}[]{0.31\textwidth}
			\centering
			\includegraphics[width=\textwidth]{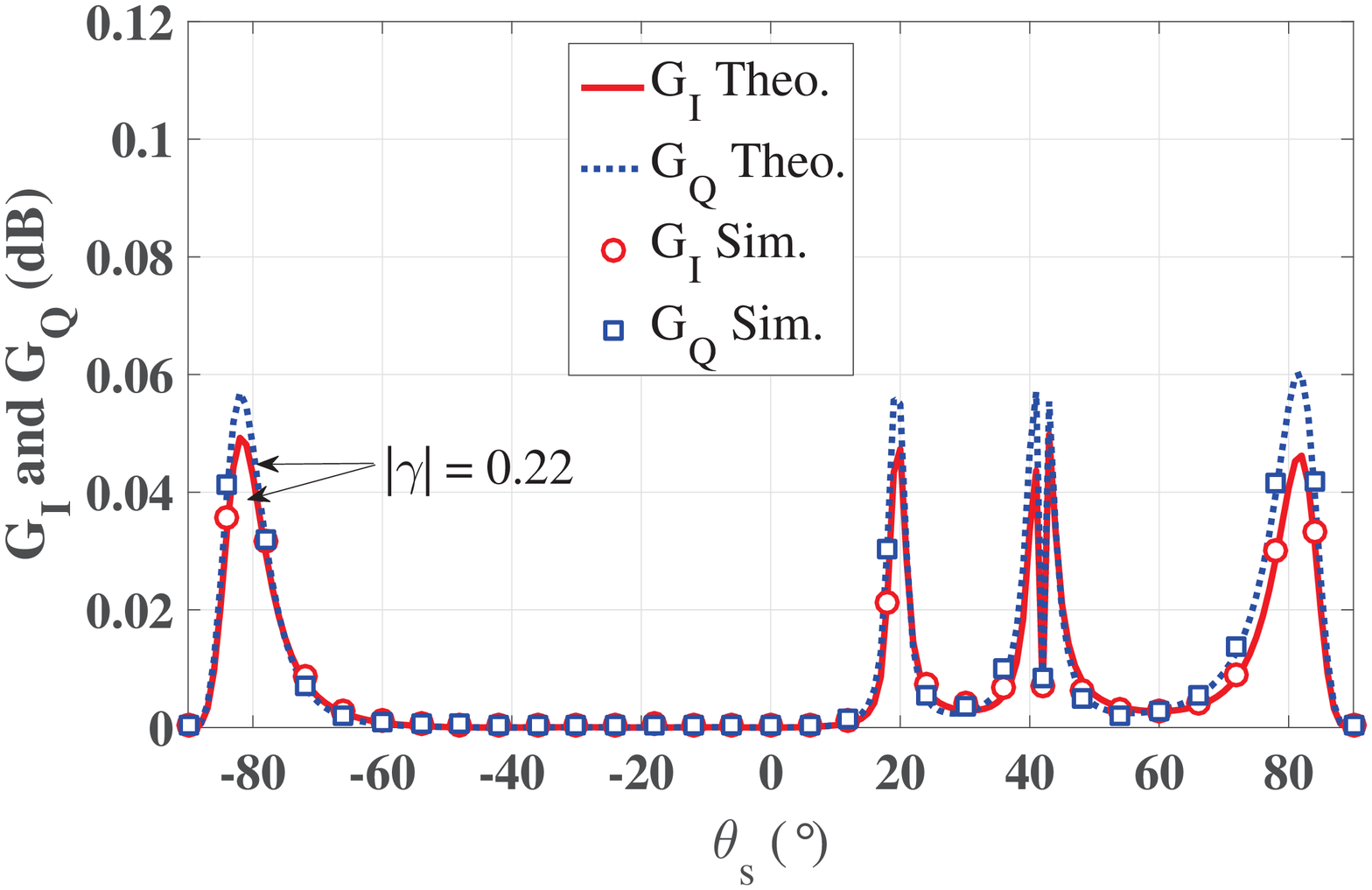}
		\end{minipage}
	}
	\subfigure[]{
	\label{fig:IQGIGQ3} 
	\begin{minipage}[]{0.31\textwidth}
		\centering
		\includegraphics[width=\textwidth]{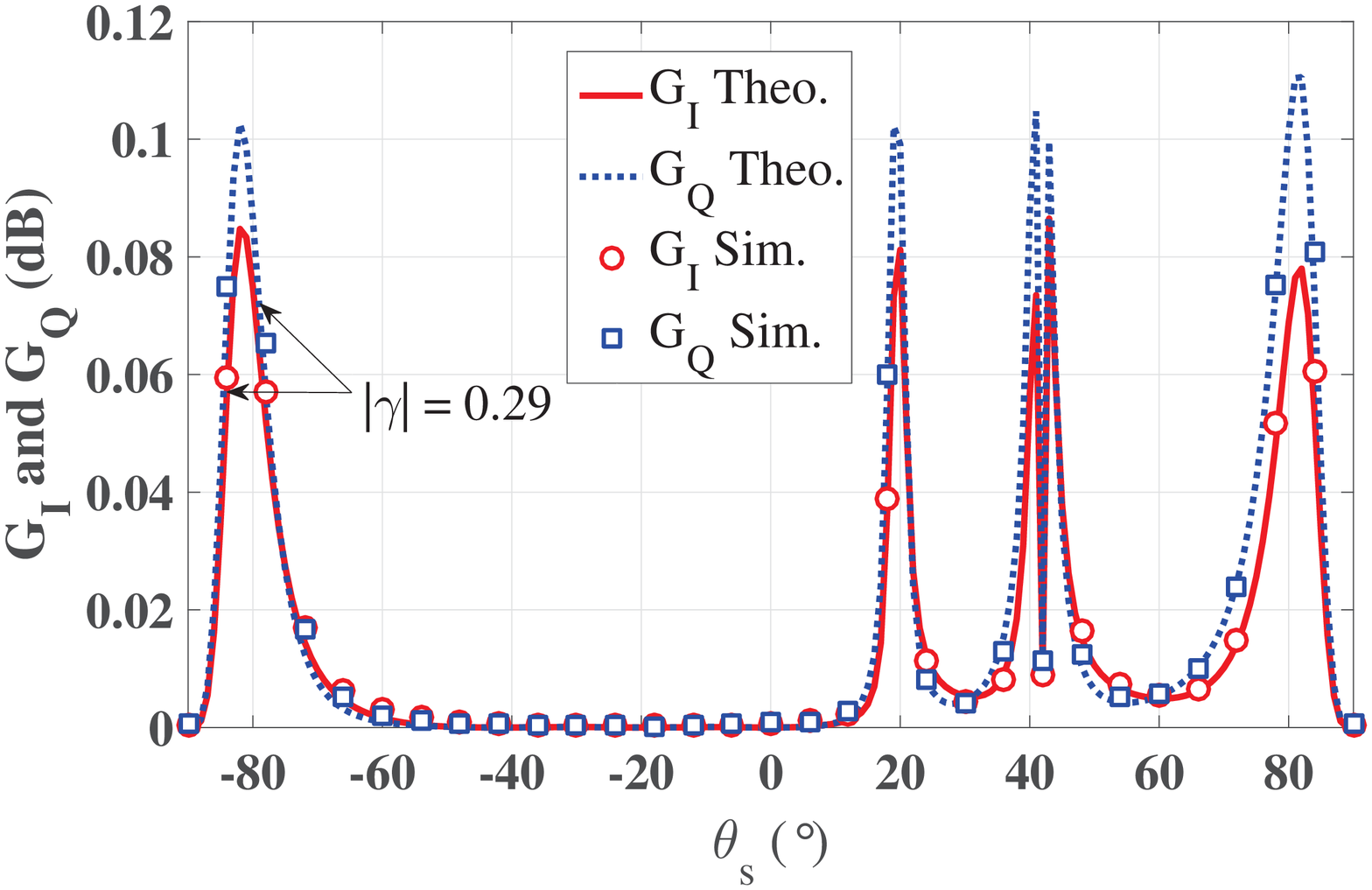}
	\end{minipage}
}
	\caption{Theoretical and simulated individual SINR gains in the I and Q channels, $G_I$ and $G_Q$, as functions of the DOA of the SOI $\theta_s$, for three types of I/Q imbalanced antenna arrays. (a) $g=1.08$, $\zeta=8.2^\circ$, (b) $g=0.9$, $\zeta=-11.2^\circ$, and (c) $g=1.15$, $\zeta=15^\circ$.}
	\label{fig:IQGIGQ} 
\end{figure*}
\begin{equation}\label{eq:SONrcoef}
\gamma_s=\gamma_p=\gamma=\frac{2\mu\nu}{|\mu|^2+|\nu|^2}.
\end{equation}
%

Conventional performance evaluation metrics in \cite{Chevalier2007,Hakkarainen2013} can only quantify the enhanced null steering capability of the WL MVDR beamformer over  the  Capon one  for  I/Q imbalanced signals in an overall way. This is shown in  Fig. \ref{fig:IQGain}, where the overall SINR gain is plotted as a function of the DOA of the SOI $\theta_s$. Three levels of amplitude and phase imbalances, ($g=1.08$, $\zeta=8.2^\circ$), ($g=0.9$, $\zeta=-11.2^\circ$), and ($g=1.15$, $\zeta=15^\circ$) were investigated, which gave $|\gamma|$ = 0.16, 0.22 and 0.29, according to \eqref{eq:SONrcoef}, respectively. The beamformer parameters were set to $N=6$, $P=3$, $\theta_1=19^\circ$, $\theta_2=42^\circ$ and $\theta_3=90^\circ$. Now, with more degrees of freedom provided by the proposed full SINR analysis framework, the individual SINR gains of the WL beamformer, $G_I$ and $G_Q$, were depicted in Fig. \ref{fig:IQGIGQ} as functions of $\theta_s$. Observe that our derived theoretical $G_I$ and $G_Q$ are in perfect agreement with their respective simulated results. Besides, enhanced attenuations of I/Q imbalance were achieved by the WL MVDR beamformer over the Capon one in both the I and Q channels, and such advantages were more pronounced for a larger SO noncircularity rate of the interferences.
\section{Conclusion}
A full second-order (SO) performance analysis framework for the WL MVDR beamformer has been introduced to provide an in-depth characterization on its statistical behavior. The full SO statistics of the output interferences and noise (IN) have made it possible to quantify individual SINR gains of the WL MVDR beamformer over the Capon one in both the in-phase (I) and quadrature (Q) channels, which are tightly connected with the overall SINR advantage via their respective \textit{distribution coefficients}. Detailed expressions of the individual SINR gains have been provided for the reception of an unknown signal corrupted by an arbitrary number of orthogonal noncircular interferences in three different spatial situations, which indicate that the advantage of the WL MVDR beamformer over the Capon one not only holds overall but also applies in both the individual I and Q channels. For rigor, we have considered the most general case to explore the link between the SO noncircularity coefficient of the input interferences and the individual SINR gains in the I and Q channels. It has been shown that the individual SINR gains in the I and Q channels are both monotonically increasing functions of the SO noncircularity rate of the interferences, and that they exhibit different increasing slopes w.r.t. the SO noncircularity phase of the interferences. Numerical simulations support the theoretical results.

\begin{appendices}
\section{Proof of the monotonically increasing nature of $G_I$ and $G_Q$ on $|\gamma|$}\label{App:AppendixA}
By comparing \eqref{eq:GI_generic_approx} and \eqref{eq:GQ_generic_approx}, we first observe that $G_I$ and $G_Q$ are symmetric functions over the coefficient $\alpha_w$, and hence we need to consider either the situation $\alpha_w>0$ or $\alpha_w<0$ only.

Now, taking $\alpha_w>0$ as an example, $\lambda_Q$ in \eqref{eq:lambdaQ_generic_approx} is a monotonically increasing function of the SO noncircularity rate $|\gamma|$ of the interferences. Recall that the distribution coefficient $\lambda_Q > 0$ is guaranteed from \textit{Remark 2}, and $G$ in \eqref{eq:G_appro} also strictly increases over $|\gamma|$. It is then straightforward to see that $G_Q$ in \eqref{eq:GQ_generic_approx} is a monotonically increasing function of $|\gamma|$. However, in the I channel, since the distribution coefficient $\lambda_I$ in \eqref{eq:lambdaI_generic_approx} is a monotonically decreasing function of $|\gamma|$, the overall effect of $|\gamma|$ on $G_I$ in \eqref{eq:GI_generic_approx} remains nonintuitive. To address this issue, we calculate the first derivative of $G_I$ w.r.t. $|\gamma|$ as
%
\begin{align}\label{dGI_approx}
	\frac{\partial G_I}{\partial |\gamma|} = \frac{\alpha_{w}|\gamma|^4-3\alpha_{w}|\gamma|^2+2|\alpha_{Is}|^2|\gamma|}{\varepsilon(1-|\alpha_{Is}|^2)(1-|\gamma|^2)^2}.
\end{align}

Note that, in the most general case, we have $0< |\alpha_{Is}| < 1$ and $0<|\gamma|<1$, so that the denominator on the right hand side of \eqref{dGI_approx} is always positive. By defining $f(|\gamma|)$ to represent the polynomials in the nominator as
\begin{equation} \label{eq:fgamma}
f(|\gamma|)\triangleq\alpha_{w}|\gamma|^4-3\alpha_{w}|\gamma|^2+2|\alpha_{Is}|^2|\gamma|,
\end{equation}
and owing to the fact that $0<\alpha_{w}\leq|\alpha_{Is}|^2$, we have
\begin{equation} \label{no}
\begin{split}
	f(|\gamma|)&\geq \alpha_{w}|\gamma|^4-3\alpha_{w}|\gamma|^2+2\alpha_{w}|\gamma| \\
	&=\alpha_{w}|\gamma|(|\gamma|+2)(|\gamma|-1)^2>0.
\end{split}
\end{equation}

Therefore, $G_I$ is also a monotonically increasing function of $|\gamma|$, although its increasing slope is more flat than that of $G_Q$ due to the decreasing monotonicity of $\lambda_I$ on $|\gamma|$.
	
\end{appendices}
	
{\small{} \bibliographystyle{IEEEtran}
\bibliography{bibliography}
 }{\small \par}

\end{document}